\def\tthdump#1{#1}
\documentclass[11pt]{article}
\input pdfdef.sty
        \usepackage{color,graphicx}
      \usepackage{lastpage}
      \usepackage{ifthen}
      \usepackage{xspace}
      \usepackage{makeidx}
      \usepackage{amsmath}
      \usepackage[pdflinkmargin=5pt,pdfstartview={FitBH -32768},pdfpagemode=None,plainpages=false]{hyperref}
      \makeindex
      
      \tthdump{\voffset= 0.0 cm}
      \tthdump{\hoffset=-2.5 cm}
      \tthdump{\setlength{\textheight}{21.5 true cm}}
      \tthdump{\setlength{\textwidth}{16.5 true cm}}
      \tthdump{\parindent=30pt}

\def\mz{\small}
\def\nz{\normalsize}
\def\lz{\large}
\def\Lz{\Large}
\def\LZ{\LARGE}
\def\LLz{\LARGE}
\def\hz{\huge}
\def\Hz{\Huge}

\newif\iftth
\def\tthdump#1{#1}
\tthdump{}  

\tthdump{}

\iftth
 
\def\cite#1{.bcit.{#1}.ecit.}

\let\ref=\cite
\else
\let\ref=\cite

\fi
\iftth

\def\ree#1{.bf.(.oah.{#1}.ncite.{#1}.qgt.\theequationppp.cas.)./bf.} 
\else

\def\ree#1{(\ref{#1})}
\fi

\iftth 
 
\else 
 
\fi 
\def\nline{
\begingroup
\iftth
\vskip 0.5cm
\else
\hskip-40cm.
\newline
\parindent=0pt
\fi
\endgroup}
\def\obcent{\begin{center}}
\def\oecent{\end{center}}
\def\obc{\begingroup\bf }
\def\oec{\endgroup}
\def\title#1{\begingroup   \nline\obcent\LLz{{\bf #1}}\oecent\tthdump{\vskip 0.5cm}\endgroup}
\def\author#1{\begingroup        \obcent\nz{ #1}\oecent\endgroup}
\def\address#1{\begingroup \nline\obcent\nz{{\em #1}}\oecent\vskip 1.0cm\endgroup}
\def\listcard{
\let\prevleng\textheight
\setlength{\textheight}{23.0 true cm}
\newpage
\setcounter{section}{97}
\listofcards
\prevleng}
\iftth
\def\listcardname{
\LARGE{\bf List of Cards}
}
\def\listofcards{\label{List of Cards}\section*{\listcardname}\input{/home/fbraga/pctex/texinput/temp/firstlochtm}}
\fi
\iftth 
\newcounter{appendico}
\newcounter{subsubsection}[subsection]
\newcounter{subsssection}[subsubsection]
   \newcounter{paragraph}[subsubsection]
\renewcommand\thesubsubsection{\thesubsection .\@arabic\c@subsubsection} 
\renewcommand\thesubsssection {\thesubsubsection.\@arabic\c@subsssection}
\renewcommand\theparagraph{Sec.\arabic{section}.\arabic{subsection}.\arabic{subsubsection}.\@arabic\c@subsssection}
\def\secto#1{\setcounter{equationp}{0}\setcounter{equation}{0} 
\section{Sec.\arabic{section} - #1}
.oa.section.\arabic{section}.ca.
.oa.sectionn.\arabic{section}.ca.
} 

\def\subsssecto#1{\addtocounter{subsssection}{1}
{\bf Sec.\arabic{section}.\arabic{subsection}.\arabic{subsubsection}.\arabic{subsssection} - #1}
.oa.section.\arabic{section}.\arabic{subsection}.\arabic{subsubsection}.\arabic{subsssection}.ca.
.oa.sectionn.\arabic{section}.\arabic{subsection}.\arabic{subsubsection}.\arabic{subsssection}.ca.
.oa.Sec.\arabic{section}.\arabic{subsection}.\arabic{subsubsection}.\arabic{subsssection}.ca.
}

\def\paragraph#1{\addtocounter{subsssection}{1}
{\bf Sec.\arabic{section}.\arabic{subsection}.\arabic{subsubsection}.\arabic{subsssection} - #1}
.oa.section.\arabic{section}.\arabic{subsection}.\arabic{subsubsection}.\arabic{subsssection}.ca.
.oa.sectionn.\arabic{section}.\arabic{subsection}.\arabic{subsubsection}.\arabic{subsssection}.ca.
.oa.Sec.\arabic{section}.\arabic{subsection}.\arabic{subsubsection}.\arabic{subsssection}.ca.
} 
\fi
\iftth          
 
\else 
 
\fi
\iftth 
\def\testnum#1#2#3{ 
\ifnum #1>#2{#3}\fi
}
\else
\def\testnum#1#2{} 
\fi 
\iftth          
\def\makfigm#1#2#3#4#5#6{  
\begin{figure}
.oa.figure.\arabic{section}f#1.ca.
.oa.figure.\Roman{appendico}f#1.ca.
\label{figure.#1}
\label{figuren.#1}
\vspace{#6}
\centerline{
\includegraphics[scale= .55]{#2}
}
\begin{card}   
       \def\thecard{Fig.#1}
\ccapl{}
          \label{figure.#1}
           \def\thecard{#1}
\ccapl{}
         \label{figuren.#1}
\end{card}
\vspace{0.55cm}
\mbox{\bf Fig.#1 #4}
\end{figure}
} 
\else 
\def\makfigm#1#2#3#4#5#6{
\begingroup
\vspace{0.5cm}
\hspace{#5}
\begin{figure}
\centerline{
\HideDisplacementBoxes
\ifpdf
\includegraphics[scale= .#3]{#2}
\else
\BoxedEPSF{#2.ps scaled #3}
\ForceHeight{10cm}
\fi
}
\vspace{#6} 
\capto{#4} 
\label{mfig #1}
\end{figure}\nulin{}} 
\fi

\def\brkn#1{\begingroup
\tthdump{\newpage\parindent=0pt \hbox to\textwidth{#1}}
\endgroup}
\def\brkl#1{\begingroup
\tthdump{\parindent=0pt \hbox to\textwidth{#1}}
\endgroup
\vskip\belowdisplayskip\noindent
}

\def\brko#1{\begingroup
\tthdump{\parindent=0pt \hbox to\textwidth{\hspace{30pt}#1}}
\endgroup}

      \def\brkm#1{\begingroup
\tthdump{\hbox to\textwidth{#1}}
\endgroup}
\def\brkk#1#2{\begingroup
\tthdump{\hbox to\textwidth{#1}\parindent=0pt \newpage{#2}}
\endgroup}
\def\brkkfirst#1#2{\begingroup
\tthdump{\hbox to\textwidth{\hspace{30pt}#1}\parindent=0pt \newpage{#2}}
\endgroup}

\def\brkkk#1#2#3{\begingroup
\tthdump{\hbox to\textwidth{#1}}
\tthdump{{#2}}
\tthdump{\parindent=0pt\newpage{#3}}
\endgroup}
\def\brkkkfirst#1#2#3{\begingroup
\tthdump{\hbox to\textwidth{\hspace{30pt}#1}}
\tthdump{{#2}}
\tthdump{\parindent=0pt\newpage{#3}}
\endgroup}
\def\brkkkf#1#2#3{\begingroup
\tthdump{\hbox to\textwidth{#1} {#2} \newpage\parindent=0pt{#3}}
\endgroup}
\iftth 

\else 
\fi 
\let\bm\bem

\def\bfr{\begin{flushright}}
\def\efr{\end{flushright}}
\iftth 
\def\bn#1{\vspace{}\vspace{}\vspace{}\beq{#1}\lef\vspace{}\vspace{}\vspace{}}
\else 
\def\bn#1{\begingroup\bfr\vspace{-0.92cm}\beq{#1}\lef\efr\vspace{-9pt}\endgroup\hspace{-4pt}}
\fi 
\iftth 
\def\see#1{\hspace{-3pt}
\setcounter{equationp}{\value{equation}}
\addtocounter{equation}{#1}
\setcounter{equationppp}{\value{equationpp}}
\addtocounter{equationpp}{#1}
.qgt.\arabic{section}.{#1}.eqp=\arabic{equationp}.\arabic{equationppp}.cas.)./bf.
\setcounter{equation}{\value{equationp}}
\setcounter{equationpp}{\value{equationppp}}
\hspace{-6pt}}
\else 
\def\see#1{\hspace{-4pt}\addtocounter{equationpp}{#1}
(\ref{\theequationpp})
\addtocounter{equationpp}{-#1}
\hspace{-6pt}$\!$}
\fi 
\def\eeq{\end{eqnarray}}
\def\eqcard#1{\vspace{#1}\begin{card}\def\thecard{\theequation}\ccap{}\tthdump{\label{\theequation}}
\end{card}}
\def\eqcards{\begin{card}\def\thecard{\theequation}\ccap{}\tthdump{\label{\theequation}}
\end{card}}
\iftth
\newcounter{equationp}[section]
\newcounter{equationpp}
\newcounter{equationppp}
\renewcommand \theequation{\arabic{section}.\arabic{equation}}

\renewcommand \theequationpp{\arabic{equationpp}}

\fi 
\iftth
\def\lefa#1{\end{eqnarray}\eqcard{#1}}
\else 
\def\lefa#1{\label{\theequationpp}\end{eqnarray}\eqcard{#1}}
\fi 
\iftth
\def\lefb{\end{eqnarray}}
\else 
\def\lefb{\label{\theequationpp}\end{eqnarray}}
\fi 
\iftth 
\def\lee#1{\end{eqnarray}~.oa.\arabic{section}.\arabic{equation}.qgt.
.oa.equation.\arabic{section}.\arabic{equation}.qgt.
.oa.{#1}.qgt.} 
\else 
\def\lee#1{\label{#1}\label{\theequationpp}\end{eqnarray}\vspace{-1.85cm}\eqcards}
\fi 
\def\lef#1{\nonumber\end{eqnarray}}
\def\lefann{\nonumber\end{eqnarray}\vspace{-0.4cm}\hspace{-5pt}}
\iftth 
\def\bee#1#2{\addtocounter{equation}{1}\addtocounter{equationpp}{1}~.oa.\arabic{section}.\arabic{equation}.qgt.
.oa.\Roman{section}.\arabic{equation}.qgt.
.oa.\arabic{equationpp}.qgt.
.oa.equation.\arabic{section}.\arabic{equation}.qgt.
\addtocounter{equation}{-1}.oa.{#1}.qgt.
.bed.
${#2}$ 
.ed1.
\addtocounter{equation}{1}(\arabic{section}.\arabic{equation})
\addtocounter{equationp}{1}\addtocounter{equationppp}{1} 
.ed2.} 
\else
\def\bee#1#2{\vspace{0.0cm}\beq{#2}\addtocounter{equationp}{1}\addtocounter{equationpp}{1}\lee{#1}}
\fi 

\def\beec#1#2#3{
\tthdump{\beq{\hspace{1.2 cm}#2}}
\tthdump{\nn\eeq\vspace{-0.6 cm}}
\tthdump{\beq{#3}}
\addtocounter{equationp}{1}\addtocounter{equationpp}{1}\lee{#1}}
\def\mathe#1{\ifmmode{#1}\else${#1}$\fi} 
\def\@empty{}

\def\nulin#1{\parindent=0pt\parskip=0pt{\tt #1}\hspace{0pt}\parindent=30pt}
\def\numat#1{\begingroup\parskip=1pt{\bf #1} \hspace{0pt}\endgroup\parindent=30pt}

\def\m#1{\ifmmode\mbox{#1}\else ${#1}$\fi}
\iftth
\def\emm#1{{\em #1}}
\else
\def\emm#1{\ifmmode\mbox{#1}\else{\em #1}\fi}
\fi
\let\embm=\emm
\iftth 
\def\mbm#1{\ifx #10\newline
\else 
{#1}
\fi} 
\else 
\def\mbm#1{\ifx #10\newline
\else 
\ifmmode\mbox{#1}\else\numat{#1}\fi
\fi} 
\fi

\def\link#1#2{
\iftth 
.ataref.{#1}.stcol..foff..bf.{#2}./bf../fon..cas.
\else 
\begingroup\htmladdnormallink{#2}{#1}\endgroup
\fi}
\def\comment#1#2#3#4#5{
\iftth
\obc\label{comment-#1}
\ifx #20
.ataref.{\$pwd/000ref/\$bwd.htm\#comment-#1}.stcol..foff..bf.\\{--------------------comment-/#1/ {#4} -------------------}\\./bf../fon..cas.\\{\em #5}
\\.ataref.{\$pwd/000ref/\$bwd.htm\#comment-#1}.stcol..foff..bf.\\{----------------------------------------------------------------------------------------------------------------------------------------------------}\\./bf../fon..cas.\\
\else 
.ataref.{\$pwd/#3\#comment-#1}.stcol..foff..bf.\\{--------------------comment-/#1/ {#4} -------------------}\\./bf../fon..cas.\\{\em #5}
\\.ataref.{\$pwd/#3\#comment-#1}.stcol..foff..bf.\\{----------------------------------------------------------------------------------------------------------------------------------------------------}\\./bf../fon..cas.\\
\fi 
\oec
\else
\ifx #20
\mbm{\obc\link{000ref/$bwd.htm\#comment-#1}{/#1/#4}\oec}
\else 
\mbm{\obc\link{#3\#comment-#1}{/#1/#4}\oec}
\fi 
\fi}

\def\jumponeline{\numat{\vspace{0.40cm}}}
\let\newline=\jumponeline
\iftth 
 
\let\noindent={.noindent.}
\else 

\fi
\iftth 
\def\vst{.noindent.} 
\def\vsm#1{.noindent.}

\def\hst{.noindent.}
\def\hsts{.noindent.}
\def\hstp{.noindent.} 
\def\hstps{.noindent.}

\else 
\def\vst{\vspace{10pt}}
\def\vsm#1{\vspace{-#1pt}}

\def\hst{\hspace{-30pt}}
\def\hsts{\hspace{-15pt}}
\def\hstp{\hspace{30pt}}
\def\hstps{\hspace{15pt}}

\fi  

\def\vbm{\begin{verbatim}}
  
\def\tcot#1#2#3{
\if #1c\if #2a{\em{#1#2#3}}\fi\fi
}
\newcommand{\tcom}[1]
{
\ifthenelse{\equal{#1}{true}}{TRUE}{}
\ifthenelse{\equal{#1}{false}}{FALSE}{}
}
\def\bfl{\begin{flushleft}}

\def\efl{\end{flushleft}}
\def\beq{\begin{eqnarray}}

\def\nn{\mathe{\qquad\qquad\nonumber}}
\tthdump{}
\iftth

\else 
\mathchardef\pmaux="2206

\fi

\def\sf{\;ba\;}

\iftth 
    \def\upo#1{.bupp.{#1}.eupp.}
     \def\up#1{.bupp.{#1}.eupp.}
\def\suo#1{.bsub.{#1}.esub.}
    \def\sub#1{.bsub.{#1}.esub.}

   \def\abs#1{¤¤0124;{#1}¤¤0124;}
\def\absu#1#2{¤¤0124;{#1}¤¤0124;\upo{#2}}
\def\abss#1#2{¤¤0124;{#1}¤¤0124;\suo{#2}}
\else 
   \def\abs#1{\ifmmode|{#1}|\else$|{#1}|$\fi}
\def\absu#1#2{\ifmmode|{#1}|^{#2}\else$|{#1}|^{#2}$\fi}
\def\abss#1#2{\ifmmode|{#1}|_{#2}\else$|{#1}|_{#2}$\fi}
\fi

                  \def\2m#1#2{{$#1{#2}$}}
                \def\3m#1#2#3{{$#1{#2#3}$}}
              \def\4m#1#2#3#4{{$#1{#2#3#4}$}}
            \def\5m#1#2#3#4#5{{$#1{#2#3#4#5}$}}
          \def\6m#1#2#3#4#5#6{{$#1{#2#3#4#5#6}$}}
        \def\7m#1#2#3#4#5#6#7{{$#1{#2#3#4#5#6#7}$}}
      \def\8m#1#2#3#4#5#6#7#8{{$#1{#2#3#4#5#6#7#8}$}}
    \def\9m#1#2#3#4#5#6#7#8#9{{$#1{#2#3#4#5#6#7#8#9}$}}
\def\lt{\m{<}}
 
\iftth

        \def\of#1{{\lz ¤¤0040;}{#1}{\lz ¤¤0041;}}

      \def\brac#1{{\lz ¤¤0040;}{#1}{\lz ¤¤0041;}}

      \def\brak#1{{\lz ¤¤0091;}{#1}{\lz ¤¤0093;}}

      \def\brar#1{{\lz ¤¤0123;}{#1}{\lz ¤¤0125;}}

\else 
      \def\brac#1{\mathe{\left({#1}\right)}}
      
        \def\of#1{\mathe{\left(#1\right)}}

   \def\brak#1{\mathe{\left[#1\right]}}

   \def\brar#1{\mathe{\left\{#1\right\}}}

\fi 

\iftth

\def\staraux{¤¤0042;}
   \def\star{¤¤0042;}
\def\circaux{¤¤2662;}
   
\else

\mathchardef\staraux="213F
        \def\star{\mathe{\staraux}}
\mathchardef\circaux="220E
        
\fi 

\iftth 

\else 

\fi 

\let\toaux=\to
\def\to{\mathe{{\toaux}}}
 
\def\lim#1{\mathe{\mathop{\tt lim}\limits_{#1}}}
\iftth                  

\def\arg{{\tt arg}}
\else 

\def\arg{\mathe{{\mbox{arg}}}}
\fi 

\iftth

\else

\fi

\let\over=\choose 

\iftth

    \def\rightarrow{{\lz ¤¤8594;}} 
\else 

\let\logaux=\log
\def\log{\mathe{\logaux}}

\let\rightarrowaux=\rightarrow 
     
    \def\rightarrow{\mathe{\rightarrowaux}}
\fi 

\iftth

      \def\da{d}
      \def\ea{{\tt e}}
      \def\fa{f}
      \def\ga{{\tt g}}
      
      \def\ia{{\tt i}}
      
      \def\ka{k}
      \def\la{l}
      \def\ma{{\tt{m}}}
      
      \def\na{n}
      
      \def\pa{p}
      \def\qa{q}
      \def\ra{r}

      \def\wa{w}
      \def\xa{x}
      \def\ya{y}
      \def\za{z}
\def\obc{\begingroup\bf }
\def\oec{\endgroup}

       \def\ac{A}
       
       \def\cc{C}
       
       \def\ec{E}

       \def\mc{M}
       \def\nc{N}
       
       \def\pc{P}

       \def\sc{S}
       
       \def\sl{s}
       \def\tc{T}
       \def\uc{U}

\else 
      
      \def\da{\mathe{d}}
      \def\ea{\mathe{{\tt e}}}
      \def\fa{\mathe{f}}
      \def\ga{\mathe{{\tt g}}}
      
      \def\ia{\mathe{{\tt i}}}
      
      \def\ka{\mathe{k}}
      \def\la{\mathe{l}}
      \def\ma{\mathe{{\tt{m}}}}
      
      \def\na{\mathe{n}}
      
      \def\pa{\mathe{p}}
      \def\qa{\mathe{q}}
      \def\ra{\mathe{r}}
      
      \def\sl{\mathe{s}}

      \def\wa{\mathe{w}}
      \def\xa{\mathe{x}}
      \def\ya{\mathe{y}}
      \def\za{\mathe{z}}

       \def\ac{\mathe{A}}
       
       \def\cc{\mathe{C}}
       
       \def\ec{\mathe{E}}

       \def\mc{\mathe{M}}
       \def\nc{\mathe{N}}
       
       \def\pc{\mathe{P}}

       \def\sc{\mathe{S}}
       
       \def\tc{\mathe{T}}
       \def\uc{\mathe{U}}

\fi  

\iftth 
    \def\min{\tt{min}}
    \def\max{\tt{max}}
   \def\prev{\tt{prev}}

\else 
 \def\min{\mathe{\tt{min}}} 
 \def\max{\mathe{\tt{max}}} 
\def\prev{\mathe{\tt{prev}}}

\fi

\iftth 
\def\stac#1#2{stacooootttttttttttt{\mz {#1}}stcstacstacstcsubfontsizemo{\nz {#2}}desubdcccccccccccccccccccc}

\else 
\def\stac#1#2{\stackrel{#1}{#2}}

\fi

\iftth 
\let\logaux={log}
\def\log{\logaux}
       \def\cdotaux={¤¤8230;} 
\def\cdots{\cdotaux}

     \def\sum={¡¡¡¡¡¡¡¡¡¡¡¡¡¡¡¡¡¡¡¡¡¡¡¡¡¡¡¡¡¡¡¡¡¡¡.bfs.{\LLz ¤¤8721;}£££££££££££££££££££££££££££££££££££££££££££££££££££££££££££££.efs.}
\def\sigsu#1#2{sssssssssssssssssssssssssssssssssss{#2}oooooooooooooooooooo{\LLz ¤¤8721;}++++++++++++++++++++++++++subfontsizemo{{#1}}desubdcccccccccccccccccccc}    

\def\sumsu#1#2{sssssssssssssssssssssssssssssssssss{#2}oooooooooooooooooooo{\LLz ¤¤8721;}++++++++++++++++++++++++++subfontsizemo{({#1})}desubdcccccccccccccccccccc}
\def\sums#1{¡¡¡¡¡¡¡¡¡¡¡¡¡¡¡¡¡¡¡¡¡¡¡¡¡¡¡¡¡¡¡¡¡¡¡.bfs.{\LLz ¤¤8721;}£££££££££££££££££££££££££££££££££££££££££££££££££££££££££££££.efs.\suo{({#1})}¢¢¢¢¢¢¢¢¢¢¢¢¢¢¢¢¢¢¢¢}
\def\sumsne#1{.bfs.{\Lz ¤¤8721;}.efs.\suo{(${#1}$)}}

\let\sumss\sumst

   \def\wau#1{{w\upo{#1}}}
\def\gt{.gt.}
\def\lt{.lt.}

\def\leq={.bfs.£.efs.}
             \def\intaux={\hz ¤¤8747;}
\def\int{iiiiiiiiiiiiiiiiiiiiiiiiiiiiiiiiiii{}oooooooooooooooooooo{\hz ¤¤8747}++++++++++++++++++++++++++subfontsizemo{}desubdcccccccccccccccccccc}
\def\intsu#1#2{iiiiiiiiiiiiiiiiiiiiiiiiiiiiiiiiiii{#2}oooooooooooooooooooo{\hz ¤¤8747}++++++++++++++++++++++++++subfontsizemo{{#1}}desubdcccccccccccccccccccc}
       \def\ointaux={\Hz ¤¤8750;}
       \def\oint{\Hz ¤¤8750;}

              \def\dointaux={\hz ¤¤8751;}
       \def\doint{\dointaux}
\def\infty={¤¤8734;} 
\def\prod={\Lz ¤¤8719;}

\let\in=\belong
\let\bel=\belong
 
 \def\foral{¤¤8704;} 

\ifmmode
\def\prodsu#1#2{sssssssssssssssssssssssssssssssssss{#2}oooooooooooooooooooo{\LLz ¤¤8719;}++++++++++++++++++++++++++subfontsizemo{({#1})}desubdcccccccccccccccccccc}
\else
\def\prodsu#1#2{sssssssssssssssssssssssssssssssssss${#2}$oooooooooooooooooooo{\LLz ¤¤8719;}++++++++++++++++++++++++++subfontsizemo{(${#1}$)}desubdcccccccccccccccccccc}
\fi

\def\prods#1{¡¡¡¡¡¡¡¡¡¡¡¡¡¡¡¡¡¡¡¡¡¡¡¡¡¡¡¡¡¡¡¡¡¡¡.bfs.{\LLz ¤¤8719;}£££££££££££££££££££££££££££££££££££££££££££££££££££££££££££££.efs.\suo{({#1})}¢¢¢¢¢¢¢¢¢¢¢¢¢¢¢¢¢¢¢¢}

   \def\aas#1{{a\suo{#1}}}
   
    \def\eu#1{{\tt e}\upo{#1}}

   \def\fas#1{f\suo{#1}}

    \def\ns#1{{n\suo{#1}}}

   \def\nus#1{.bfs.n.efs.\suo{#1}}
    \def\nusi{.bfs.n.efs.\suo{i}}

   \def\ras#1{r\suo{#1}}
   
   \def\wau#1{w\upo{#1}}

    \def\xu#1{x\upo{#1}}

   \def\xas#1{x\suo{#1}}
   \def\xau#1{x\upo{#1}}

    \def\yu#1{y\upo{#1}}

   \def\yas#1{y\suo{#1}}
   \def\yau#1{y\upo{#1}}

\else 
              \def\cdotaux{\mathinner{\cdotp\cdotp\cdotp}}
\def\cdots{\mathe{\cdotaux}} 

\mathchardef\sumaux="1350
       \def\sum{\mathe{\sumaux}}
 \def\sigsu#1#2{\mathe{\sumaux_{{#1}}^{#2}}}

 \def\sumsu#1#2{\mathe{\sumaux_{({#1})}^{#2}}}
    \def\sums#1{\mathe{\sumaux_{({#1})}}}
    \def\sumsne#1{\mathe{\sumaux_{({#1})}}}
    
\def\sumss#1#2{\mathe{\sumaux_{{#1}\over{#2}}}}

\def\leq{\mathe{\leqaux}} 
\let\intaux=\int 
          \def\int{\mathe{\intaux}}

    \def\intsu#1#2{\mathe{\intaux_{{#1}}^{#2}}} 
\let\ointaux=\oint 
         \def\oint{\mathe{\ointaux}} 
        \def\doint{\mathe{\ointaux\ointaux}}

\mathchardef\inftaux="0231
\def\infty{\mathe{\inftaux}}
\def\prodsu#1#2{\mathe{\prod_{#1}^{#2}}}

   \def\prods#1{\mathe{\prod_{#1}}}
   
               \let\inaux=\in
    \def\in{\mathe{\,\inaux}\,}
   \def\bel{\mathe{\,\inaux\,}}

               \let\foraux=\forall 

 \def\foral{\mathe{\foraux}}

   \def\aas#1{\mathe{a_{#1}}}
   
    \def\eu#1{\mathe{{\tt e}^{#1}}}

   \def\fas#1{\mathe{f_{#1}}}

    \def\ns#1{\mathe{n_{#1}}}

   \def\nus#1{\mathe{\nuaux_{#1}}}
   
    \def\nusi{\mathe{\nuaux_{i}}}

   \def\ras#1{\mathe{r_{#1}}}
   
   \def\up#1{\mathe{^{#1}}}
   \def\sub#1{\mathe{_{#1}}}

   \def\wau#1{\mathe{w^{#1}}}

    \def\xu#1{\mathe{x^{#1}}}

   \def\xas#1{\mathe{x_{#1}}}
   \def\xau#1{\mathe{x^{#1}}}

    \def\yu#1{\mathe{y^{#1}}}

   \def\yas#1{\mathe{y_{#1}}}
   \def\yau#1{\mathe{y^{#1}}}

\fi

\iftth 

     \def\cap{c¤¤0039;}

     \def\gap{g¤¤0039;}

\else 

     \def\cap{\mathe{c'}}

     \def\gap{\mathe{g'}}

\fi

\iftth 

    \def\ecsk{E\suo{k}}

\else 

    \def\ecsk{\mathe{E_{k}}}

\fi

\iftth                  

   \def\ecs#1{E.bsub.{#1}.esub.}

   \def\ucs#1{U.bsub.{#1}.esub.}

\else 

   \def\ecs#1{\mathe{E_{#1}}}

   \def\ucs#1{\mathe{U_{#1}}}

\fi 

\let\hbaux=\hbar
          \def\hbar{\mathe{\hbaux\hspace{1pt}}}

\let\lamba=\lambdabar
     \def\lambdabar{\mathe{\lamba}}
        \let\muaux=\mu
        \let\nuaux=\nu
        
\iftth 
\let\otimx={¤¤8855;}
       \def\otimes{\otimx} 
\def\approx{¤¤8776;}
\det\app{¤¤8776;}

\def\propaux{¤¤8733;}

        \def\b{¤¤0946;}
       \def\be{¤¤0946;}

      \def\del{{\lz ¤¤8706;}.efs.}
        
       \def\de{¤¤0948;}

        \def\e{¤¤1297;}
       \def\ep{¤¤1297;}
    \def\eps#1{¤¤1297;\suo{#1}}

  \def\lambaux{¤¤0955;}

   \def\lambda{\lambaux}

       \def\mu{¤¤0956;}

       \def\nu{¤¤0957;}
        
    \def\piaux{¤¤0960;}
        
       \def\pi{¤¤0960;}
       \def\Pi{¤¤0928;}
      
      \def\rho{¤¤0961;}
        
      \def\chi{¤¤0962;}
       \def\xi{¤¤0958;}
       \def\Xi{¤¤0926;}
   \def\chiaux{¤¤0962;}
    \def\xiaux{¤¤0958;}
    \def\Xiaux{¤¤0926;}

       \def\om{¤¤1120;}
      \def\ome{¤¤1120;}
    
      \def\psi{¤¤1137;}
   
\else 
\mathchardef\xiaux="0118
\mathchardef\chiaux="011F
\mathchardef\Xiaux="7004
\mathchardef\geaux="3215 
\mathchardef\rhoaux="011A
\mathchardef\psaux="0120
\mathchardef\Psaux="7009
\mathchardef\omegaux="0121
\mathchardef\Omegaux="700A
\mathchardef\leaux="3214 
\let\gtaux=>
\let\ltaux=<
\let\otimx=\otimes
       \def\otimes{\mathe{\otimx}}
\mathchardef\propaux="322F

\mathchardef\approxaux="3219
        \def\approx{\mathe{\;\approxaux\;}}
        \let\app=\approx
         \let\lambaux=\lambda

   \def\lambda{\mathe{\lambaux}}

      \def\del{\mathe{\partial}}

    \def\eps#1{\mathe{\epsilon_{#1}}}

\def\gt{\mathe{\gtaux}}
\def\lt{\mathe{\ltaux}}

\def\leq{\mathe{\leaux}}

\def\b{\mathe{\beta}}

\def\e{\mathe{\epsilon}}

\def\rho{\mathe{\rhoaux}}

\let\piaux=\pi
\let\piauxc=\Pi

\def\be{\mathe{\beta}}
\def\de{\mathe{\delta}}
\def\ep{\mathe{\epsilon}}
\def\mu{\mathe{\muaux}}
\def\nu{\mathe{\nuaux}}
\def\om{\mathe{\omegaux}}
\def\ome{\mathe{\omegaux}}
        \def\Omega{\mathe{\Omegaux}}

        \let\omc=\Omega
        
\def\pi{\mathe{\piaux}}
\def\Pi{\mathe{\piauxc}}

\def\psi{\mathe{\psaux}}
\def\kapp{\mathe{\kappa}} 

\def\chi{\mathe{\chiaux}}
\def\xi{\mathe{\xiaux}}
\def\Xi{\mathe{\Xiaux}}
\fi

\iftth
\mathchardef\capux="225C
\mathchardef\cap="225C
\mathchardef\intersec="225C
\mathchardef\intersection="225C
\mathchardef\cupux="225B
\mathchardef\cup="225B
\mathchardef\union="225B
\let\fracux=\mathe{\frac}
\else 
\mathchardef\capux="225C
\def\cap{\mathe{\capux}}
\let\intersec=\cap
\let\intersection=\cap
\let\fracux=\frac
\mathchardef\cupux="225B
\def\cup{\mathe{\cupux}}
\def\frac#1#2{\mathe{\fracux{#1}{#2}}}
\fi

\let\union=\cup
\iftth
\def\fracs#1#2{
\ifmmode\frac{#1}{#2}\else\mathe{{#1}\left/{#2}\right.}\fi} 
\else 
\def\fracs#1#2{
\ifmmode\fracux{#1}{#2}\else\mathe{\hspace{-3pt}{#1}\hspace{-4pt}\left/{#2}\right.}\fi} 
\fi 
 
\iftth

      \def\cald{\lz .bf.{D}./bf..efs.}
      
      \def\caln{\lz .bf.{N}./bf..efs.}
      \def\calr{\lz .bf.{R}./bf..efs.}

      \def\caly{\lz .bf.{Y}./bf..efs.}

  \def\calds#1{{\lz .bf.{D}./bf..efs.}_{#1}}

  \def\calns#1{{\lz .bf.{N}./bf..efs.}_{#1}}

   \def\caldsk{{\lz .bf.{D}./bf..efs.}_{k}}
   \def\calnsk{{\lz .bf.{N}./bf..efs.}_{k}} 
    
\else

\def\cald{\mathe{\begingroup \cal{D}\endgroup}} 
 
\def\caln{\mathe{\begingroup \cal{N}\endgroup}} 
\def\calr{\mathe{\begingroup \cal{R}\endgroup}}

\def\caly{\mathe{\begingroup \cal{Y}\endgroup}}

\def\calds#1{\mathe{{\begingroup \cal{D}\endgroup}_{#1}}}

\def\calns#1{\mathe{{\begingroup \cal{N}\endgroup}_{#1}}}

\def\caldsk{\mathe{{\begingroup \cal{D}\endgroup}_{k}}} 
\def\calnsk{\mathe{{\begingroup \cal{N}\endgroup}_{k}}} 
 
\fi 

\iftth

\else

\fi

\tthdump{   }
\tthdump{   }
\tthdump{   }
\tthdump{   }
            
\tthdump{   }
\tthdump{   }
\tthdump{   }
\tthdump{   }
\tthdump{   }

            \def\ripl-ii{RIPL-2}
            \def\ripl{RIPL-2}
            \def\ripl2{RIPL-2}

            \def\ripl-iis{{RIPL-2} }
            
            \def\ripl2s{{RIPL-2} }

            \def\ripl-iib{{RIPL-2} }
            
            \def\ripl2b{{RIPL-2} }
\def\capit#1{
\if #1=a A\fi
\if #1=b B\fi
\if #1=c C\fi
\if #1=d D\fi
\if #1=e E\fi
\if #1=f F\fi
\if #1=g G\fi
\if #1=h H\fi
\if #1=i I\fi
\if #1=k K\fi
\if #1=j J\fi
\if #1=l L\fi
\if #1=m M\fi
\if #1=n N\fi
\if #1=o O\fi
\if #1=p P\fi 
\if #1=q Q\fi
\if #1=r R\fi
\if #1=s S\fi
\if #1=t T\fi
\if #1=u U\fi
\if #1=v V\fi
\if #1=w W\fi
\if #1=x X\fi
\if #1=y Y\fi
\if #1=z Z\fi
} 

\tthdump{}

\def\nse#1#2#3{{\em Nucl. Sci. Eng. }{\bf  #1}, #2 (#3)}

\tthdump{}
\tthdump{}
\tthdump{}
\tthdump{}
\tthdump{}
\tthdump{}
\tthdump{}
\tthdump{}
\tthdump{}
\tthdump{}
\tthdump{}

\begin{document}

  \pagestyle{myheadings}
  \markright{\thepage}


\vspace{-3.5cm} 
\title{\bf Brief critical analysis of the  Darwin-Fowler method}
\vspace{-1.0cm} 
\author{F. B. Guimaraes}
\vspace{-1.7cm} 
\address{
Instituto de Estudos Avan\c cados/DCTA,\\ 
12228-001 S\~ao Jos\'e dos Campos, S\~ao Paulo, Brazil \\ 
e-mail: fbraga@ieav.cta.br} 
\vspace{-1.0cm} 
\begin{abstract} \mz{\embm{\hsts We present a brief numerical study of the  Darwin-Fowler
method applied to the analysis of the energy partition of essembles of bosons and
fermions. 
We analyze the assertion of the existence of a ``strong maximum" made in the original
paper of Darwin and Fowler and other studies and show that although the presumed saddle
point along the real axis of the grand canonical parameters may exist it cannot, in
general, be characterized as ``strong", in the sense of having much larger magnitude than
the other points along the path of integration.\\
In addition, we show that in some cases the saddle point is not even present and the
various approximations of the method can be interpreted as a tricky reformulation of
usual thermodynamic relations.\\ 
The close connection of the method with the formalism of the Laplace transform may
produce wrong results if the internal energy of the components of the ensemble is not
large enough.\\ 
Therefore, although useful in many applications the Darwin-Fowler method may not be
suitable, in general, for a detailed microscopic analysis of the nuclear structure in
connection with the Shell Model approach, as it is usually done in studies of the
pre-equilibrium stage of nuclear reactions.} 
} 
\end{abstract} 


 

\parindent=30pt
\pagenumbering{arabic}

\secto{Introduction}

The seminal paper of Darwin and Fowler\cite{df22} proposed a method for the statistical determination
of how the energy is partitioned in an ensemble of a large number of microsystems, which should be used
to replace the traditional approach based on the direct computation of probabilities. The
statistical analysis of these ensembles showed that the ``most probable arrangement" would have a much
greater probability than the others and their effort intended to avoid the use of the Stirling's
approximation for factorials, which they recognized as ``illegitimate" in many cases. 

In nuclear physics the method of Ref.\cite{df22} is often used as an auxiliary formalism in the
description of the pre-equilibrium stage of nuclear reactions (PE) and the nuclear density in connection with
the Shell Model and the exciton model (EXM).\cite{b68,w71,e60,k72} 

In particular, it can be used with the traditional Shell Model approach to define the
moments of the Hamiltonian in terms of Laplace transforms and their inverses to obtain the nuclear
density\cite{w71} and in the microscopic description of PE dynamics 
to obtain the transition strengths.\cite{obp}  

In the formalism of Ref.\cite{obp} some problems were observed in connection with the Darwin-Fowler method,
due to the statistical nature of its approximations, which 
may not be very precisely defined at the microscopic level and can obscure 
the analysis of some details of the microscopic interaction in the description of the PE
process.\cite{fbgarXiv1} 

In the Darwin-Fowler formalism the level density can be defined as the pole of the grand
canonical generating function \fa\brac{\xa,\ya} divided by adequate factors {\xau{\ac+1}\yau{\ec+1}}, where
\xa\ is a parameter associated with the total number of single particle levels (sp-levels), \ac, and
\ya\ is a parameter associated with the total energy, \ec. The sp-levels are supposed to have well
defined energies, which can be degenerate or not.  

Either for bosons or fermions \fa\brac{\xa,\ya} can be expanded as a sum of products of terms of the type
(\xa\yau{\nus{\ia}}), where \nus{\ia} is an integer. Then, \fa\brac{\xa,\ya} can be interpreted, in the 
continuous approximation limit (\emm{CAP}), as the inverse Laplace transform of a linear combination of
nuclear densities for given \ac\ and \ec, which is very useful in practical
applications.\cite{b68,w71,obp} 


For this description to be physically meaningful the terms of the expansion must 
decrease in modulus when {\ya} vary over complex circles around the origin in comparison with its value at
the positive real axis, and this point of maximum should also be a minimum along the positive real
axis.\cite{df22,b68} 
Therefore one must have a \embm{saddle point} located on the positive real axes of \ya.  

In Ref.\cite{df22} the existence of the saddle point is then 
considered in connection with the method of \embm{steepest descents}. The formalism uses a \embm{qualitative
analysis} of the generating function of the ensemble to conclude that, for large \ec, this point is also a
``strong maximum" along the direction of the path of integration, taken to be the circle centered at the
origin with radius equal to the abscissa on the positive real axis where the minimum occurs.

Reference \cite{df22} then establishes that, if the integrand has a saddle point with these characteristics,
this would be sufficient to obtain approximate equations of state in the usual form for a system of bososn,
fermions, etc., for given {\ac} and {\ec}. 

In this paper, we briefly analyze the assertion of the existence of this ``strong maximum" for bosons and
fermions 
and show that although the saddle point may exist the maximum cannot be characterized in general as ``strong", in
the sense of having much larger magnitude than the other points along the path of integration.

In \ref{section.2} we review some basic definitions of the Darwin-Fowler statistics, the Shell Model
formalism 
and the definition of the level density to better explain the details
of the arguments presented in this Introduction.

In \ref{section.3} we present the simplest cases of partition functions that can be considered as
physically meaningful, following closely the development of Ref.\cite{df22}, and show that the
``strong maximum" hypothesis does not necessarily holds for these functions. 
Similar results are obtained in \ref{section.4} for more realistic functions of ensembles with fixed number
of ``particles" and non degenerate levels, in the fundamental state or for non null excitation. The influence of
the statistical parameter associated with the chemical potential is also analyzed. 

At last, we present in \ref{section.5} a quick review of the use of the Laplace transform in connection with
the Darwin-Fowler method, an explanation of why the approximated eqations obtained with the
method are correct for various applications and the general conclusion.

\secto{The level density of a system of bosons or fermions}

The analysis of Ref.\cite{df22} is focused on a set of ``Planck vibrators" (PV), obeying the Bose-Einstein
statistics, with a given energy distribution, i.  e., \aas{\ra} vibrators with energy \ep\ra, for
fixed energy unit \ep\ and variable integer \ra, and the statistical ensemble is defined by the
constants for the total energy, \ec, and the total number of vibrators, \mc, satisfying
\bm{ \sums{\ra}\aas{\ra}=\mc, \mbox{\hstp and \hstp} \sums{\ra}\ra\ep\aas{\ra}=\ec 
\;. }  
Due to the indistinguishability of the PV with the same energy, the number of different sets satisfying
the first equation in \see{-0} is given by 
\bm{ \frac{\mc!}{\aas{0}!\aas{1}!\aas{2}! ... }  \; }  
and the second equation constrains the possible sets \brar{\aas{0},\aas{1},\aas{2} ... } so that their total
number could be symbolically written as 
\bm{ \cc = \sumss{\aas{0},\aas{1},\aas{2} ... }{\mc,\ec=fixed}\frac{\mc!}{\aas{0}!\aas{1}!\aas{2}! ... }  \;, }  
which Ref.\cite{df22} calls the total number of ``complexions" representing the ensemble, for given \mc\ and
\ec. Now, if one considers the following series
\bm{  (1+\xau{ \ep}+ \xau{2\ep}+ \cdots )\up{\mc} = (1-\xau{\ep})\up{-\mc}  \; }  
in which \mc\ is supposed to be a finite number, possibly very large, then the multinomial expansion gives 
\bm{ \sums{\aas{0}+\aas{1}+\aas{2}+ ... =\mc} \frac{\mc!}{\aas{0}!\aas{1}!\aas{2}! ... } 1\up{\aas{0}}
(\xau{\ep})\up{\aas{1}} (\xau{2\ep})\up{\aas{2}} \cdots = \sums{\aas{0}+\aas{1}+\aas{2}+ ... =\mc} \frac{\mc!}{\aas{0}!\aas{1}!\aas{2}! ... } 
\xau{\sums{\ra} \ra\ep\aas{\ra} } \; } 
and, therefore, the first equation of \ree{2.1} is always satisfied. Then, Eq.\see{-1} corresponds
to the statistical description of an ensemble with fixed number of PV and variable energy, given by
\sumsne{\ra}\ra\ep\aas{\ra}, and \cc\ can also be written as the sum of the coefficients of the
expansion \see{-0} in which the exponent of \xa\ satisfies the energy equation in \ree{2.1}. 

Notice that \xa\ is an arbitrary algebraic parameter with no specific physical meaning so far and one may
also eventually let \mc\ vary, to obtain the usual grand canonical description with variable number of
microsystems and variable energy. 
 

Then, equation \ree{2.3} can be rewritten using Cauchy's theorem applied over function \ree{2.4} along a path
around the origin of the \xa-complex plane with \abs{\xa}\lt 1,  
\bm{ \cc= \frac{ 1}{2\pi\ia} \oint \frac{\da\xa}{\xau{\ec+1} (1-\xau{\ep})\up{\mc} } \;, }
and one may interpret \cc\ as the total degeneracy of the PV with energy \ec. 

To obtain an approximate solution of \ree{2.6} Ref.\cite{df22} makes use of the  \embm{steepest
descents} method and consider that for \xa\ along the positive axis, the integrand becomes infinite at
\xa=0 and \xa=1, therefore it has at least one minimum in the interval, \abs{\xa}\bel(0,1) with
\arg(\xa)=0. A direct computation shows that this minimum is in fact unique and situated at a point
\xi\bel\calr, 0 \lt\ \xi\ \lt 1, and the integration path (contour) can be taken as the circumference centered
at the origin with radius \xi. Then, Ref.\cite{df22} states that ``for values of \xa\ on the contour,
\xa=\xi\ corresponds to a \embm{strong maximum}" for which the ``whole value of the integral is contributed
by the contour in the neighborhood of this point". 

This important statement is one of the essential aspects of the entire method, but it is presented as
self-evident and without further proof.


The bosonic formalism outlined above is analogous to the usual statistical description of the nucleus
(many fermions system) inspired by the Shell Model, in which the nuclear grand canonical
ensemble, for all energies (\ec) and mass numbers (\ac), can be defined by the following relations\cite{b68} 
\bee{det1}{ \ac=\sums{i}\ns{i} \;, }
where \ns{i}\in\brar{0,1}, are all possible occupation numbers of the single particle (sp)
states associated with the corresponding set of sp-levels with energies \bee{det2}{ \epsilon_i = \nusi
\epsilon \;.}
and total nuclear energy given by
\bee{det3}{ E = \caln \ep =  \sums{i} n_{i} \nus{i} \ep \;.} 
Here \nusi\ are integers and \ep\ is usually considered as an arbitrarily fixed real number,
defining the approximate ``equidistant spacing" between any two consecutive sp-levels or an ``average
spacing" of more realistic bases for the sp-states as, e. g., the H.O. basis, etc. 
 
In this context, the nuclear level density at the energy \m{E}, \m{\rho(E)}, can be {\em defined} as the
ratio between the nuclear degeneracy and the sp-level spacing \m{\ep}, using a formal description
based on the Darwin-Fowler method.\cite{b68,df22}  In this description, the generating function of the grand
canonical ensemble is given by the following expression 
\bm{ f(x,y)= \prods{i}\brac{1+\xa\yau{\nus{i}}}= \prods{i}\brac{1+\xas{i}}  
\;,  } 
where \xa\ and \ya\ are independent parameters associated with \ac\ and \ec, respectively, and the last
simpler form takes into account the fact that \xa\ is the same for all sp-states.\cite{fbgarXiv1} 

  In the grand canonical ensemble, describing a statistical set of many-body systems with variable number of
  microsystems and variable energy, the parameter \xa\ has a more strictly combinatorial meaning while \ya\ is
  related with the probability distribution associated with the various component systems of the
  ensemble. 

Then, the nuclear level density can be directly defined as an adequate {\em
pole} of the generating function divided by \m{\ep},\cite{b68}
\bee{det6}{ \rho(\ac,\ec) = \frac{1}{\brac{2\pi i}^2 \ep}
\doint\frac{f(x,y)\da\xa\da\ya}{\xu{A+1}\yu{\caln + 1}} \;, }
therefore, a first drawback in the formalism of \cite{df22} when applied to nuclear systems, although a not
necessarily an important one, is that 
expressions like \ree{2.11} for the nuclear density can only be considered as an exact definition if
\xa\ and \ya\ are continuous variables, which is equivalent to assume the continuous approximation
limit (\emm{CAP}).\cite{fbgarXiv1}

The generating function in \see{-1} can be rewritten as  
\bee{det16}{
f(x,y) = 1 + x \sums{j} y^{\nus{J}} + x^2 \sums{j1, j2} y^{\brac{\nus{J1}+\nus{J2}}} + \cdots
+ x^A \sums{j1, \cdots, jA} y^{\brac{\nus{J1}+ \cdots + \nus{JA}}} + \cdots \;, 
} 
which, therefore, describes all nuclear systems with all possible ``mass numbers" and energies (nuclear
temperature).
In other words, for a given nuclear system each \embm{configuration of sp-states} is also a microstate of the
canonical ensemble with fixed mass number and temperature,\cite{reif} and the term proportional to \xau{\ac}
is the sum over all possible configurations with fixed nuclear mass \ac\ and variable energy.  

For each nuclear level, \m{\ecsk}, corresponds usually many different configurations of sp-states and to each
nuclear mass \m{A} a term, \m{\caly_A}, is defined in Eq.\see{0} as follows,  
\bee{det20}{
\caly_A = \sums{j1, \cdots, jA} y^{\brac{\nus{j1} + \cdots + \nus{jA}}}
\;.}
which can be rewritten in terms of the degeneracies for the various nuclear levels, \m{\caldsk}, as
\bee{det21}{ \caly_A = \sums{k} \caldsk y^{\calnsk} }

\hst where
\bee{det22}{ \calnsk = \sigsu{i=1}{A} \nus{ki} = \frac{\ecsk}{\ep} }
and \m{\caldsk = \caldsk(\ecsk, A)} is the degeneracy of the nuclear level
\m{\ecsk}, for a given nuclear mass number \m{A}.   

Taking Eq.\see{-1} into Eq.\see{-3}   
and using the definition \ree{det6} for the nuclear level density yields 
\bee{det240}{
\rho(\ecsk, A) = \frac{1}{\ep} \caldsk(\ecsk, A)\;,}
which is a natural result in a microscopic description.\cite{fbgarXiv1} 
Therefore, equation \see{-4} can be rewritten as
\beec{det24}{
  f(x,y) = 1 + x   \sums{k1} \calds{k1}(\ecs{k1},1) y^{\calns{k1}(1)} + \nn}{
               x^2 \sums{k2} \calds{k2}(\ecs{k2},2) y^{\calns{k2}(2)} +
      \cdots + x^A \sums{k } \calds{k }(\ecs{k },A) y^{\calns{k }(A)} + \cdots + }
where different indices have been used for each term to reinforce the fact
that the corresponding nuclear levels may not be the same. In these expressions the sum over \ka\ is
equivalent to the sum over the nuclear energy \ecs\ka\ and \fa(\xa,\ya) can be written as a sum over nuclear
energies 
\bee{det230}{ f(x,y) = \sums{A,\uc} \cald(\ac,\uc)\xau{A}\yau{\uc/\ep} \;,}
and also as a sum over \embm{individual configurations}, with all degneracies are equal one, 
\bee{det230}{ f(x,y) = \sums{conf} \xau{A} \yau{\uc/\ep} \;.}

\secto{The simplest partition functions}

The simplest cases of physically meaningful partition functions corresponding to the description of an
ensemble of degenerate fermions or bosons can be directly calculated algebraically and show the main 
problems with the analysis of Refs.  \cite{df22} and \cite{b68}.

The essential part of the integrand of the generating function of the ensemble of identical bosons analyzed by
\cite{df22} can be written as 
\bm{ \fa(\ya)= \frac{1}{\yau{\ec+1}(1-\ya)\up{\mc}}  \;, }
which obviously approaches infinite for real positive values of \ya\rightarrow~1\sub{-} and
\ya\rightarrow~0\sub{+}. A more general definition similar to the integrand of \ree{2.11} would be
\bm{ f(x,y)= \frac{1}{\xau{\mc+1}\yau{\ec+1}} \prods{i} \frac{1}{(1-\xa\yau{\nus{i}})} 
  \;. }
Equations \see{-1} or \see{-0} represent a set of \mc\ ``Planck vibrators" (bosons) with total energy \ec. In
particular, \see{-1} is closer to the expression analyzed in \cite{df22} with each vibrator possessing the
same set of possible energies 
\pc=\brar{\na\ep;\na=0,\cdots,\infty},
where \na\ is an integer \ep\ is the energy of the basic vibrating mode. In a sligthly more general case than
\see{-1} and still using only one statistical parameter, one could associate a different basic energy for each
boson to obtain 
\bm{ \fa(\yas{1},...,\yas{\mc}) =\frac{1}{\wau{\ec+1}}\prodsu{\ia=1}{\mc}
\frac{1}{(1-\yas{\ia})} \;, } 
where \yas{\ia}=\wau{\eps{\ia}}, \wa=\eu{-\be} and  
\bm{\ec=\sumsu{\la=1}{\mc} \ras{\la}\eps{\la} \mbox{ ; \;\;} \ras{\la}\bel \{0,1,\cdots,\infty\} \;.}
 
Therefore, the simpler description at \see{-3} can be obtained by setting \eps{\ia}=\ep=fixed, \foral\ia, 
and it contains the essential numerical characteristics of the more general case \see{-1}, allowing 
\see{-3} to be used to analyze the general analytical behavior of \see{-1}. 

The above simple expressions for \fa(\ya) imply that it has at least one minimum in the interval (0,1)
which can be determined analytically 
\bm{ \frac{\da}{\da\ya}\brak{\yau{-(\ec+1)}
(1-\ya)\up{-\mc}}= \frac{1}{\yau{\ec+1}(1-\yau{\mc})} \brac{ \frac{-(\ec+1)}{\ya} +
\frac{\mc}{(1-\ya)} } = 0 \;, }  
then
\bm{ \yas{\min} = \frac{\ec+1}{\mc+\ec+1} }
and the minimum is unique. Now, following Ref.\cite{df22}, we need to analyze what happens when one
considers the values of \fa\of\ya\ for \ya\ belonging to the circumference centered at the origin of the
complex \ya-plane with radius \yas{\min}. In the qualitative analysis of \cite{df22} it is suggested that
\fa\of{\yas{\min}} would correspond to a ``strong maximum" in comparison with the other points of the
circumference, but the following simple analysis shows that this conclusion is not necessarily true. 

For example, consider a simple proportionality 
\bm{ \ec=\mc/\ka \;, }
to analyze the behavior of \see{-1} for large \mc. Then, 
\bm{ \yas{\min} = \frac{\mc+\ka}{\mc(\ka+1)+\ka} \;, }
and the point of minimum along the real axis will also be \embm{the maximum} (\fas{\max}) along the
circumference of radius \yas{\min} centered at the origin, giving 
\bm{ (\yas{\min}\up{\ec+1})\fas{\max} = (1-\yas{\min})\up{-\mc} =
\brak{\frac{\mc\ka}{(\mc(\ka+1)+\ka)}}\up{-\mc} =
\brak{\frac{\mc(\ka+1)+\ka}{\mc\ka}}\up{\mc} \;, }
and the minimum along the circumference, \fas{\min}, is located at the point \ya\ where (1-\ya) is maximum,
corresponding to \arg(\ya)=\pi\ or \ya=-\yas{\min}, is given by 
\bm{ (-\yas{\min})\up{\ec+1}\fas{\min} = (1+\yas{\min})\up{-\mc} =
\brak{\frac{\mc(\ka+2)+2\ka}{\mc(\ka+1)+\ka}}\up{-\mc} =
\brak{\frac{\mc(\ka+1)+\ka}{\mc(\ka+2)+2\ka}}\up{\mc} \;. }
  Now, one can estimate how ``strong" the maximum along the real axis is by calculating the ratio,
  \rho=\abs{\fas{\max}/\fas{\min}}, which gives 
\bm{\rho= \brac{\frac{\ka+2}{\ka}+\frac{2}{\mc}}\up{\mc} =
\brac{\frac{\ka+2}{\ka}}\up{\mc} \brac{1+\frac{(2\ka/\ka+2)}{\mc}}\up{\mc} \;, }
then,
\bm{ \lim{\mc\to\infty} \brac{\frac{\fas{\max}}{\fas{\min}}} = \brac{\frac{\ka+2}{\ka}}\up{\mc} 
\eu{\brac{2\ka/\ka+2}} \;,}
which for any \ka\gt 1 would produce \brac{\frac{\fas{\max}}{\fas{\min}}}\to\infty\ for   
\mc\to\infty, in agreement with the qualitative reasoning of \cite{df22}. 

On the other hand, one may also take \ec\ \embm{fixed} and  \ka\ varying with \mc\ in \ree{3.6}, to
analyze the behavior of the subsets consisting of the microcanonical ensembles with \ec\ and \mc\
fixed, or systems with energy non greater than an arbitrarily fixed maximum. 
In this case the first term of \see{-1} becomes 
\bm{ \pa\of{\mc,\ec} = \brac{\frac{\ka+2}{\ka}}\up{\mc} = \brac{\frac{(\mc+2\ec)/\ec}{\mc/\ec}}\up{\mc} 
= \brac{1+\frac{2\ec}{\mc}}\up{\mc}
\;,}
giving 
\bm{\lim{\mc\to\infty} \pa\of{\mc,\ec} = \eu{2\ec}  \;, }
while the second term gives
\bm{ \qa\of{\mc,\ec} = \eu{2\ka/\ka+2} = \eu{2\mc/\mc+2\ec} = \eu{[2/1+2(\ec/\mc)]}
\stac{\mc\to\infty}{=} \eu{2} \;, }
then, the total expression in \see{-4} yields
\bm{\lim{\mc\to\infty} \pa\of{\mc,\ec}\qa\of{\mc,\ec} = \eu{2 (\ec+1)}  \;, }
which can be ``very large" \embm{or not}, depending on \ec. 

One may ask also how these results translate for the analysis of an ensemble of \embm{fermions}, as in the
study of Ref.\cite{b68}, in which the number of particles is also variable and the energies of the
fermions cannot be considered the same in general. In this case we have to deal with various levels
with different energies and a generating function given by \ree{2.10}, in which we shall neglect for a
while the term ``\xa" associated with the chemical potential. Then, the generating function can be rewritten as 
\bm{ f(y)= \prods{i}\brac{1+\yau{\nus{i}}} }
where the \nus{i} are different integers in general for each \ia, characterizing the a different single
particle level for each fermion. Each term in the parenthesis can be compared with the expansion of
the terms 1/(1-\yas{\ia}) in \see{-14} and the absence of the corresponding terms proportional to
\yau{ 2\nus{i}}, \yau{ 3\nus{i}}, etc., is an expression of the exclusion principle.  

For \ya=\eu{-\be\ep}\lt 1, the various terms in \yau{\nus{i}} contribute less to the magnitude of \fa\of{\ya} 
than if they were replaced by \ya\ only.  Therefore, to analyze the asymptotic behavior of the magnitude of
\fa\of{\ya} for a ``large number" of fermions one may consider all the \nus{i} as equal to 1, which will
imply that the magnitude of the generating function is necessarily smaller than that of the analyzed
\fa\of{\ya}. 
In general, this simplification is \embm{not} physically possible for an actual system of
fermions due to the exclusion principle, unless all them belong to the same set of degenerate sp-states. 
 
We then make this approximation and analyze the following simpler expression for the integrand of \ree{2.11} 
\bm{ f(y) = \frac{ 1}{\yau{(\ec+1)}} \prodsu{i}{\mc}(1+\ya) = \frac{ (1+\ya)\up{\mc}}{\yau{(\ec+1)}} 
\;,}
 A more general definition similar to \ree{2.10} and \ree{2.11} would be
\bm{ f(x,y)= \frac{1}{\xau{\mc+1}\wau{\ec+1}} \prodsu{ i}{\mc} (1+\xa\yas{i}) 
  \;. }
where \yas{\ia}=\wau{\eps{\ia}}, \wa=\eu{-\be} and  
\bm{\ec=\sumsu{\la=1}{\mc} \ras{\la}\eps{\la} \mbox{ ; \;\;} \ras{\la}\bel \{0,1\} \;.}

 Using the interpretation \ya=\eu{-\be\ep}, equation \see{-2} would correspond to a
set of degenerate fermions with energy \ep, while \see{-1} describes fermions with different energies, for
example, \eps{i}=\ep\nus{i} with \nus{ i} integer.

  From \see{-2} one readly conclude that, for \ya\ belonging to the positive real axis and \mc\gt(\ec+1),
  \fa\of{\ya} is an incresing function of \ya\ and it is also divergent for \ya\to 0. Then, it has at least
  one minimum for \ya\gt 0, which can be determined analytically, 
\bm{ \fa'(\ya) = \frac{ \da}{\da\ya} \brak{ \frac{(1+\ya)\up{\mc}}{\yau{(\ec+1)}}}  =
\brac{ \frac{ (1+\ya)\up{\mc-1}\yau{\ec} }{ \yau{(2\ec+2)} } } (\mc\ya-(\ec+1)(1+\ya)) = 0 
\;,}
giving 
\bm{\yas{\min}= \frac{(\ec+1)}{\mc-(\ec+1)} \;. }
Then, \yas{ \min} is unique and \fa\of{ \yas{ \min}} will also be the maximum along the circumference
centered at the origin with radius \yas{ \min}. Taking as before \ec=\mc/\ka\,, results 
\bm{ \fas{ \max} = \frac{
\brak{1+(\mc+\ka)/(\mc(\ka-1)-\ka)}\up{\mc}}{ (\mc+\ka)/(\mc(\ka-1)-\ka)} = \frac{
1}{[\mc(\ka-1)-\ka]\up{\mc-1}} \brac{ \frac{ (\ka\mc)\up{ \mc}}{\mc+\ka}} \;, }
and the corresponding minimum along this circumference is 
\bm{\fas{\min} = \fa\of{-\abs{\yas{ \min}}} = 
\frac{ (\mc\ka-2\mc-2\ka)\up{ \mc}}{ (\mc+\ka)[\mc(\ka-1)-\ka]\up{\mc-1} } 
\;, }
then,
\bm{ \abs{\frac{\fas{\max}}{\fas{\min}}}  = \brac{ \frac{\mc\ka}{\mc\ka-2\mc-2\ka} }\up{\mc} =
\brac{ \frac{ \ka}{\ka-2}}\up{\mc} \brak{ \frac{ 1}{[1- 2\ka/\mc(\ka-2) ]\up{\mc} }} 
\;, }  
and for \ka\ fixed and greater than 2 results 
\bm{ \lim{\mc\to\infty} \abs{\frac{\fas{\max}}{\fas{\min}}} = \brac{\frac{\ka}{\ka-2}}\up{\mc} 
\eu{\brac{2\ka/\ka-2}} 
\;,}
which is very similar to \ree{3.9}. Now, taking \ec=\mc/\ka= constant yields
\bm{ \frac{\ka}{\ka-2} = \frac{\mc}{\mc-2\ec} \;,}
and \see{-2} becomes 
\bm{ \abs{\frac{\fas{\max}}{\fas{\min}}} = 
\brac{ \frac{\mc}{\mc-2\ec}}\up{\mc} \brak{ \frac{ 1}{[1- 2/(\mc-2\ec)]\up{\mc} }} = \brac{
\frac{1}{1-2\ec/\mc}}\up{\mc}  \frac{ \mc(1-\frac{2\ec}{\mc})\up{\mc}}{ \mc(1-\frac{2\ec+2}{\mc})\up{\mc}} 
\;, }   
then, 
\bm{ \lim{\mc\to\infty} \abs{\frac{\fas{\max}}{\fas{\min}}} = \eu{2(\ec+1)} 
\;,}
which coincides with \ree{3.13} and is not necessarily a ``very large" ratio.
  Notice that in the general case one should consider \nus{i} greater than 1, which would
  produce even smaller ratios in this case.

\newpage
\makfigm{1}{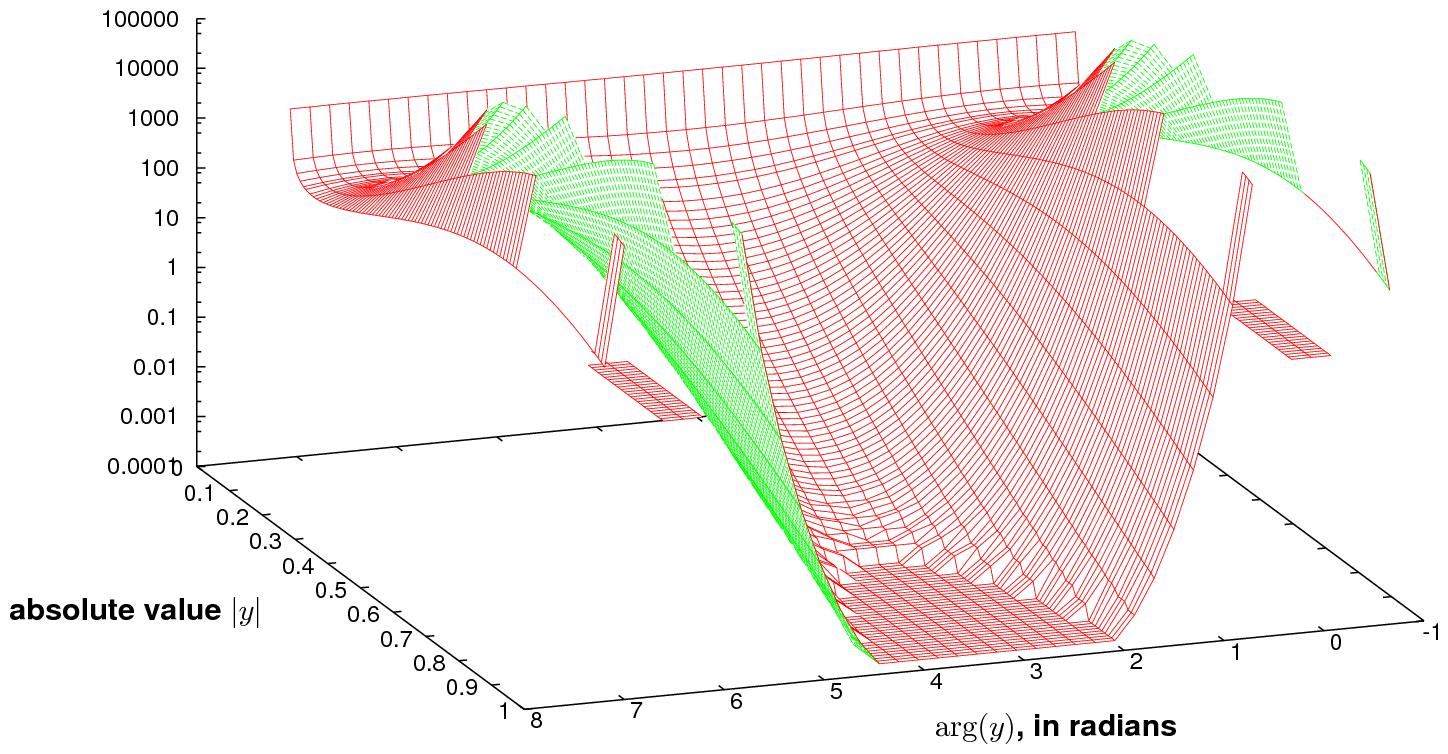}{335}{Simplified generating function for bosons \fa(\ya) of Eq.(3.1),
for variable \ya\ in the case of \mc=20 sp-states.}{-1.3cm}{-2.5cm} 

\makfigm{3}{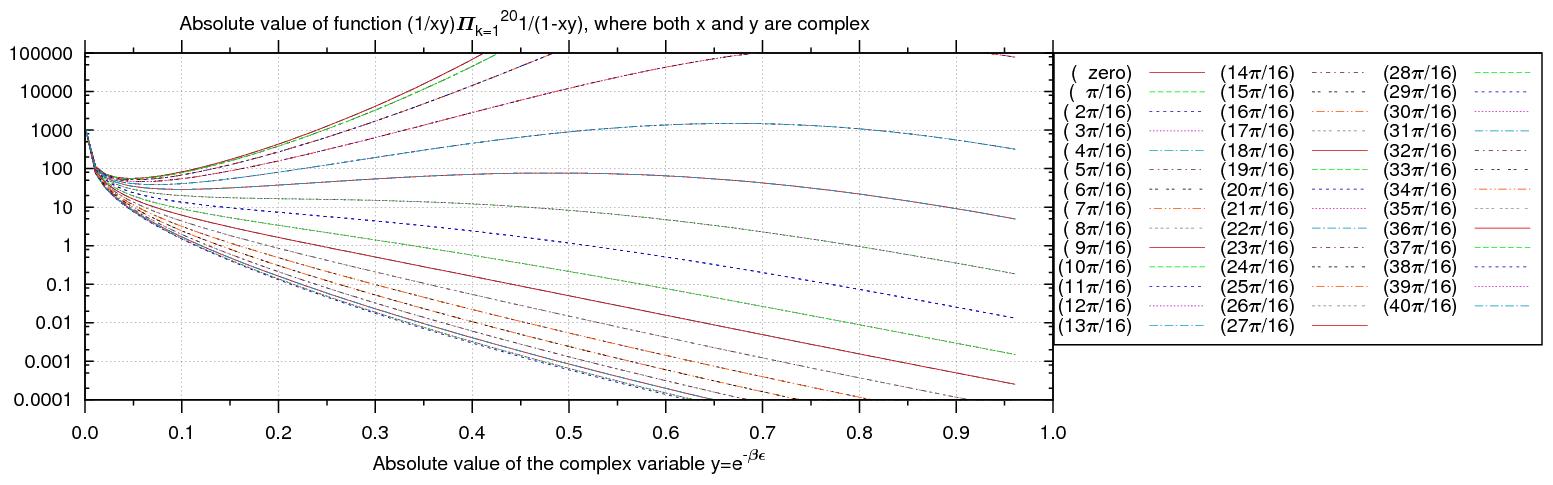}{345}{Projection of the graph of Fig. 1 on the plane containing \abs{\ya} and the
\za-axis. The various values of arg(\ya) are highlighted with different colors}{-0.6cm}{-2.5cm}

\newpage
\makfigm{2}{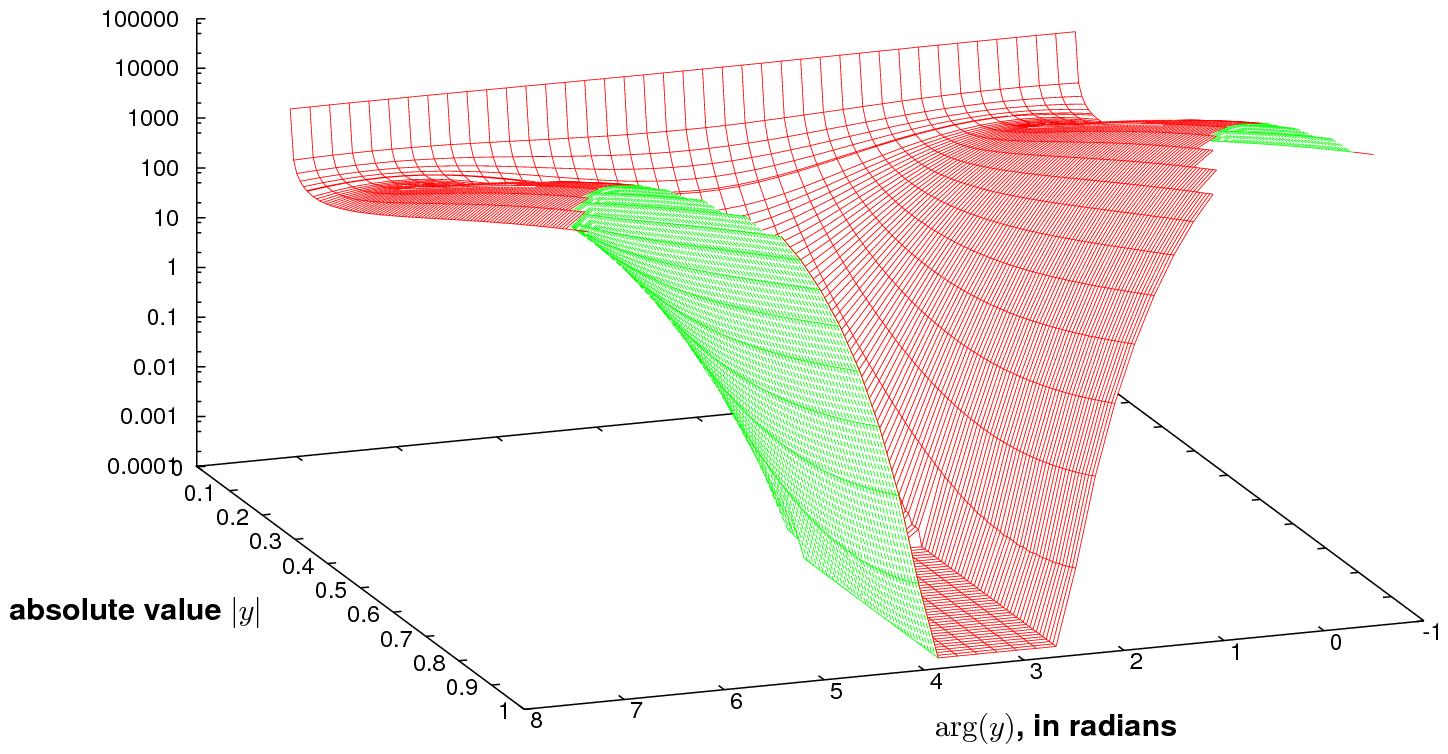}{335}{Simplified generating function for fermions \fa(\ya) of Eq.(3.17), for
variable \ya\ in the case of \mc=20 sp-states.}{-1.3cm}{-2.5cm} 

\makfigm{4}{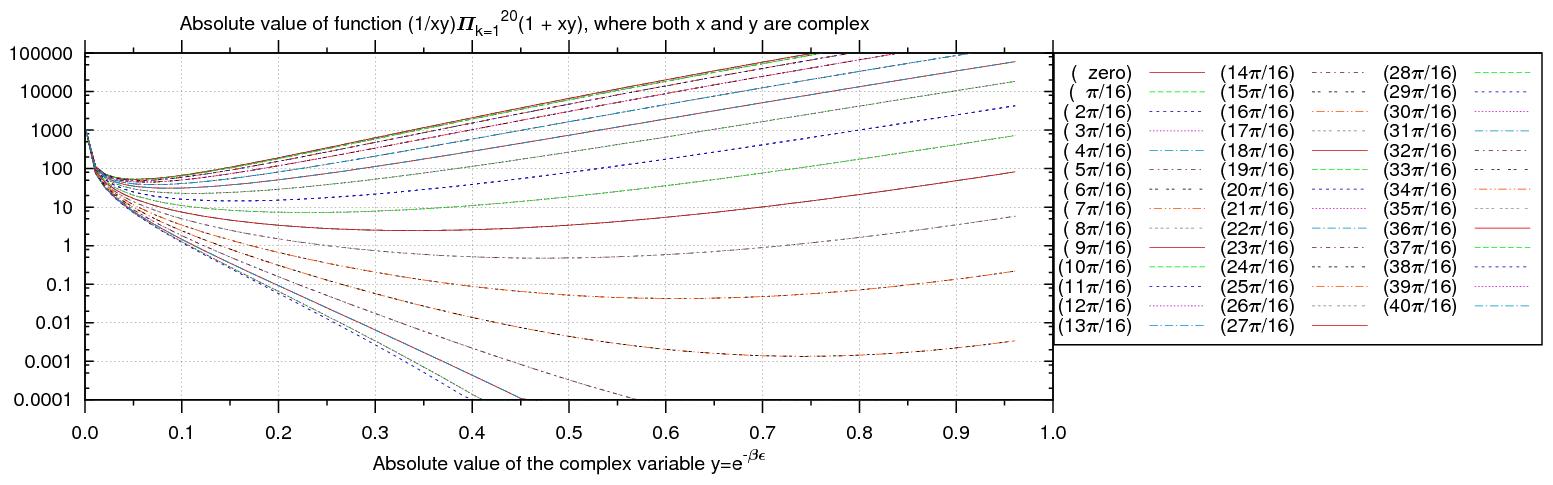}{345}{Projection of the graph of Fig. 2 on the plane containing
\abs{\ya} and the \za-axis. The various values of arg(\ya) are highlighted with different 
colors.}{-0.6cm}{-2.5cm}

\newpage
Figures \ref{figuren.1} and \ref{figuren.3} show the simplified generating function for bosons and
fermions, corresponding to Eqs. \ree{3.1} and \ree{3.17} respectively, for variable \ya\ in the case of 
\mc=20 ``particles" (``bosons" or ``fermions") and \ec=0, the fundamental state. The minimum along the
real axis as given by Eqs. \ree{3.5} and \ree{3.19}, can be seen as a broader white band crossing the
graph parallel to the arg(\ya) axis, close to the points with \abs{\ya}=0. It is clear in both graphs
that the minimum along the real axis is a saddle point, but it is \embm{not} a strong maximum along
the band.

Figures \ref{figuren.2} and \ref{figuren.4} show the corresponding projections of the three dimensional graphs 
on the plane containing the \abs{\ya} axis and the \za-axis. In these graphs it becomes evident that the
maximum along the real axis for \ya=\yas{\min} is no more than 10 times greater than the minimum along the
circumference containing \yas{\min}, in agreement with \ree{3.16} and \ree{3.29}.

Therefore, for the simple approximate partition functions considered in this section, either for
bosons or for fermions, the assumption of ``strong maximum" along the direction of the contour, in
equations like \ree{2.6} or \ree{2.11}, cannot be considered as generally valid as it is assumed in the
method of Ref.\cite{df22}.

\secto{The saddle point problem} 

A saddle point of a function is one that is stationary but not a local
extremum. For functions of two variables, it is a maximum for the variation of one
variable and a minimum for the variation of the other. More precisely, it is a point 
(\xa\sub{\star},\ya\sub{\star})\bel\calr\up{\na+\ma} that satisfies\cite{benzi}
\bm{L(\xa\sub{\star}, y) \;\leq\; L(\xa\sub{\star},\ya\sub{\star}) \;\leq\; L(x,\ya\sub{\star})
\;\mbm{,\hspace{0.5cm}}\foral x \bel \calr\up{\na} \mbm{ and }\foral y \bel \calr\up{\ma} \;,  
}   

\hst or, equivalently, 
\bm{ \mbox{min}\sub{\xa}\mbox{max}\sub{\ya} L(x, y) = L(\xa\sub{\star},\ya\sub{\star}) = 
\mbox{max}\sub{\ya}\mbox{min}\sub{\xa} L(x,y)\;,}
and the definition for complex variables would correspond to the dimensions \na=\ma=2 for two variables, or
\na=\ma=1 along perpendicular directions for one variable.  


In a first stage of the formal application of the method of Ref.\cite{df22} for partition functions, either for
fermions in the integrand of Eq.\ree{2.11} or bosons in \ree{3.2}, the variables \xa\ and \ya\ can be thought as
possessing no direct physical meaning and to have been created only to keep track of the counting of the
number of particles and the energy of the nuclear levels. As we saw in \ref{section.2} this tracking is
formally performed using the Cauchy's Theorem in the definition of the "density" associated with \ya\ for a
given \ac\ (number of particles, excitons, Planck vibrators, etc.).

For given \ya\ the analysis of the dependence of \fa(\xa,\ya) with \xa\ in \ree{2.10} is similar to
the ``simplest cases" of \ref{section.3} and essentially reduces to an additional phase on the \ya\ term.
Therefore, one can limit the analysis to the dependence on \ya\ to have an idea of the general behavior of
the partition function.

In a general expression like \ree{3.2} or \ree{3.18} one cannot deduce algebraically the point of
maximum in a straightforward way as happened in \ree{3.5} and \ree{3.21}, but due to the smooth
analytical behavior of these functions one can rely on numerical procedures to determine this point.

These functions can be easily calculated numerically by fixing one variable, for example fixed \abs{\xa} and
 \arg(\xa), and varying \abs{\ya} with values smaller than one (in accordance to its physical interpretation
 as \ya=\eu{-\b\e}, with \b\ and \e\ real and positive) and variable \arg(\ya) with a grid of points dense
 enough and reasonable range, for example between 0 and 2\pi, with steps of 2\pi/16. 

We used the Newton-Raphson method\cite{Newtonr} to obtain the maximum along the real axis of \ya\ 
and compared it with the values in other directions with the same
\abs{\ya}. 
The results are presented in the following figures and are analogous to the simpler cases of \ref{section.3}, i.
e., although the saddle point along the real axis is well defined the maximum cannot be considered as ``very
strong" in the sense of Ref.\cite{df22}. 

\makfigm{5}{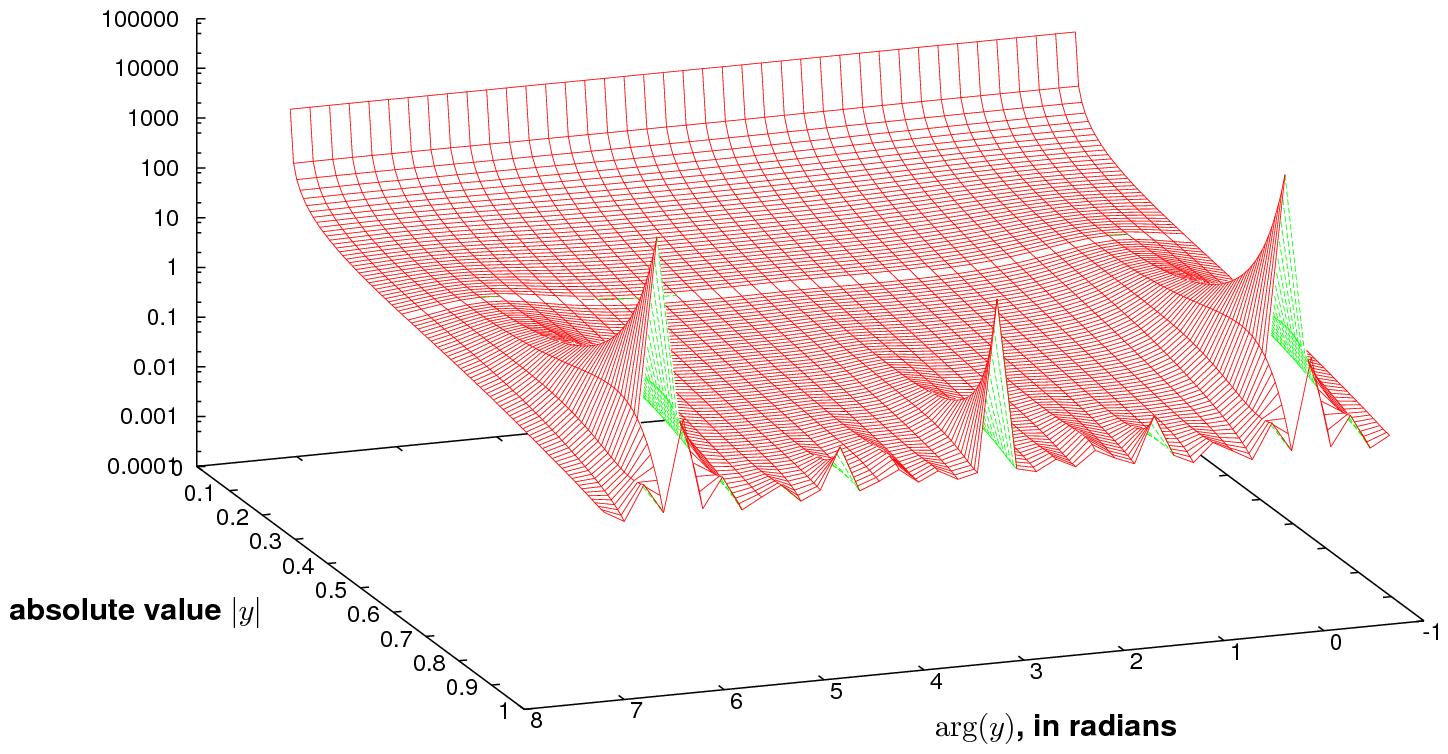}{335}{Simplified generating function for bosons \fa(\yas{1},...,\yas{\mc}) of Eq.(3.3) in
the case of \mc=20 sp-states, with \ec=0 and \eps{\la}=\la\ in (3.4).}{-1.3cm}{-2.5cm}

\makfigm{6}{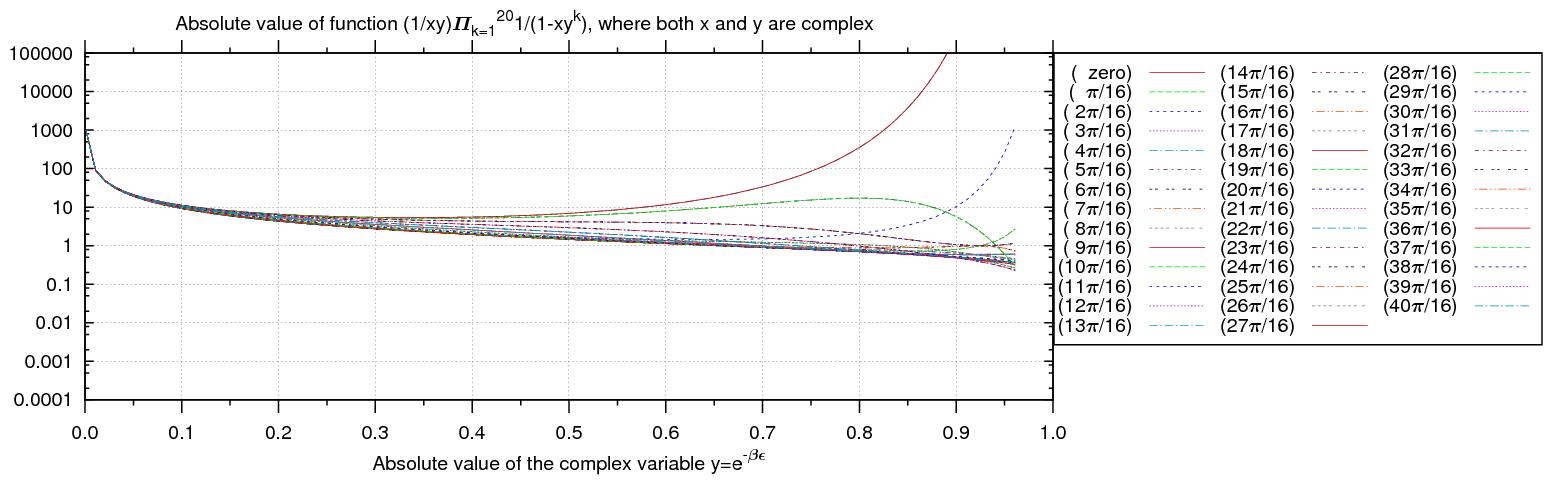}{345}{Projection of the graph of Fig. 5 on the plane containing \abs{\ya} and the
\za-axis. The various values of arg(\ya) are highlighted with different colors}{-0.6cm}{-2.5cm}

\makfigm{7}{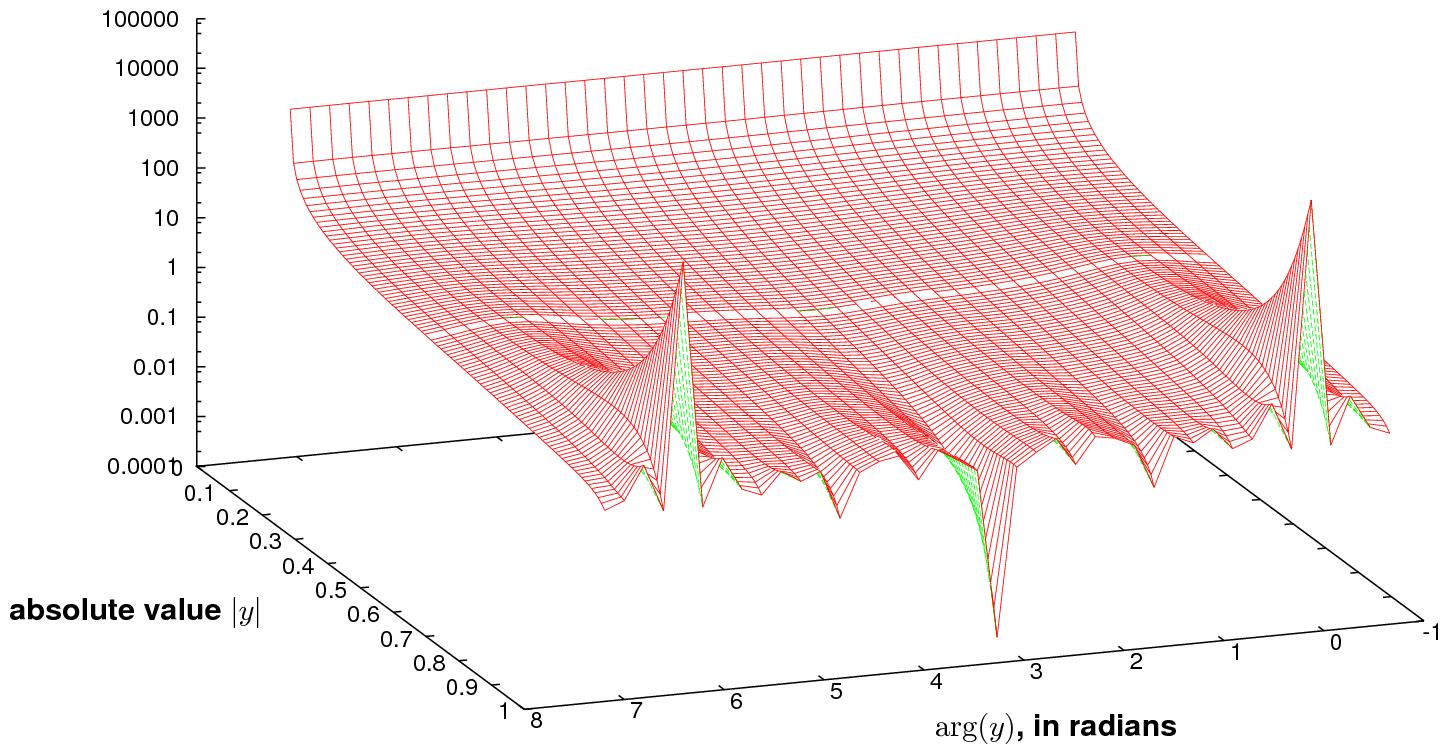}{335}{Simplified generating function for fermions \fa(\xa,\ya) of Eq.(3.19),
in the case of \mc=20 sp-states and \ec=0 and \eps{\la}=\la\ in (3.20).}{-1.3cm}{-2.5cm}

\vst 
\vst  
\makfigm{8}{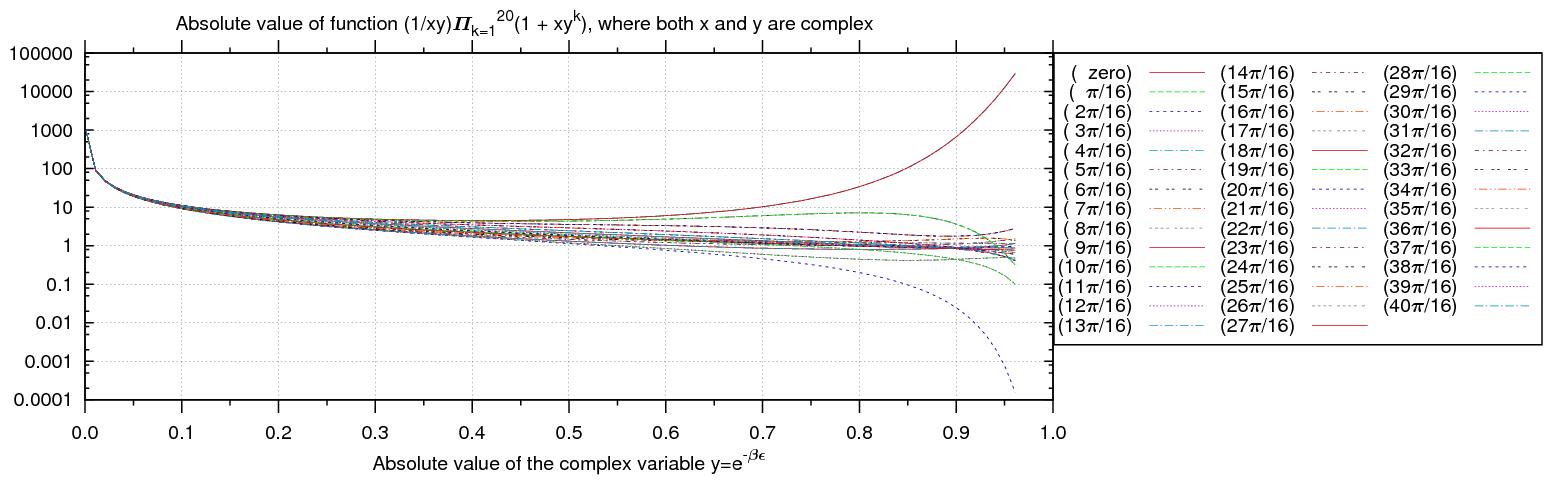}{345}{Projection of the graph of Fig. 7 on the plane containing
\abs{\ya} and the \za-axis. The various values of arg(\ya) are highlighted with different
colors.}{-0.6cm}{-2.5cm}

As expected, from the reasoning of the previous section, the maximum along the real axis results ``weaker" than
in the simplified cases because the magnitude of \ya\ is less than 1.  This is reflected by 
\embm{flatter} surfaces obtained now in comparison with the previous section, as we can see in figures
\ref{figuren.5} and \ref{figuren.7}.  These figures show the simplified partition functions for bosons and
fermions, corresponding to Eqs.\ree{3.3} and \ree{3.19}, for \mc=20 ``particles" in the fundamental state,
using arg(\xa)=0. 

The minimum along the real axis 
can be seen as a broader white band crossing the graph approximately at its middle, parallel to the
arg(\ya) axis. It is clear in both graphs that the minimum along the real axis is also a saddle point,
but it is not a strong maximum along the 
band as a function of arg(\ya).

Figures \ref{figuren.6} and \ref{figuren.8} show the projections, corresponding to figures \ref{figuren.5}
and \ref{figuren.7} respectively,  on the plane containing the \abs{\ya} axis and the \za-axis. These
graphs clearly show that the maximum along the real axis for \ya=\yas{\min} has an even smaller ratio to
other points along the band than the cases considered in the previous section, which is expected due to the
higher powers of \ya\ involved in the present functions.

Figures \ref{figuren.9} and \ref{figuren.11} show the partition functions for bosons and fermions,
corresponding to Eqs.\ree{3.2} and \ree{3.19}, for \mc=20 ``particles" in the fundamental state, using
now arg(\xa)=7\pi/8. 

\makfigm{9}{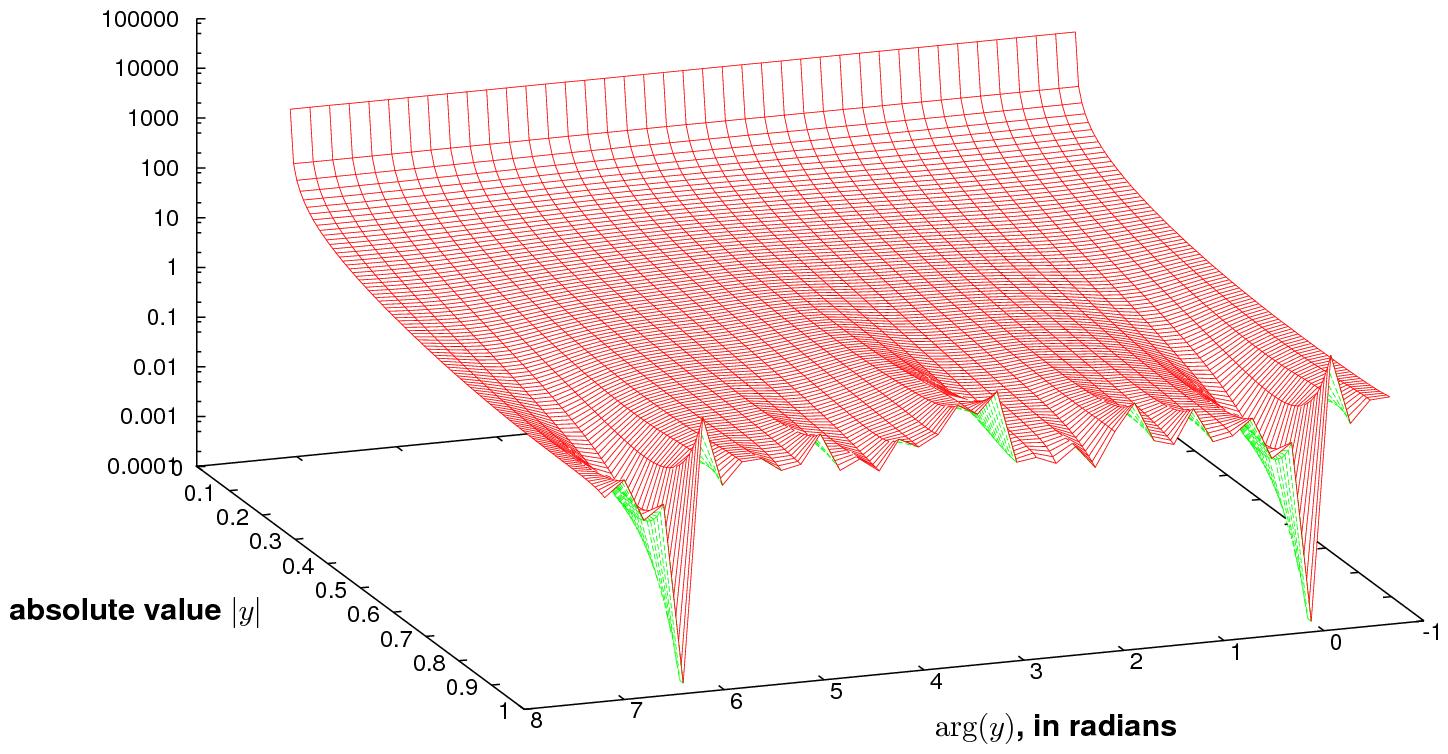}{335}{Simplified generating function for bosons \fa(\xa,\ya) of
Eq.(3.2), for variable \ya\ and fixed \xa, with arg(\xa)=7\pi/8, in the case of \mc=20
sp-states, in the fundamental state.}{-1.3cm}{-2.5cm} 

\makfigm{10}{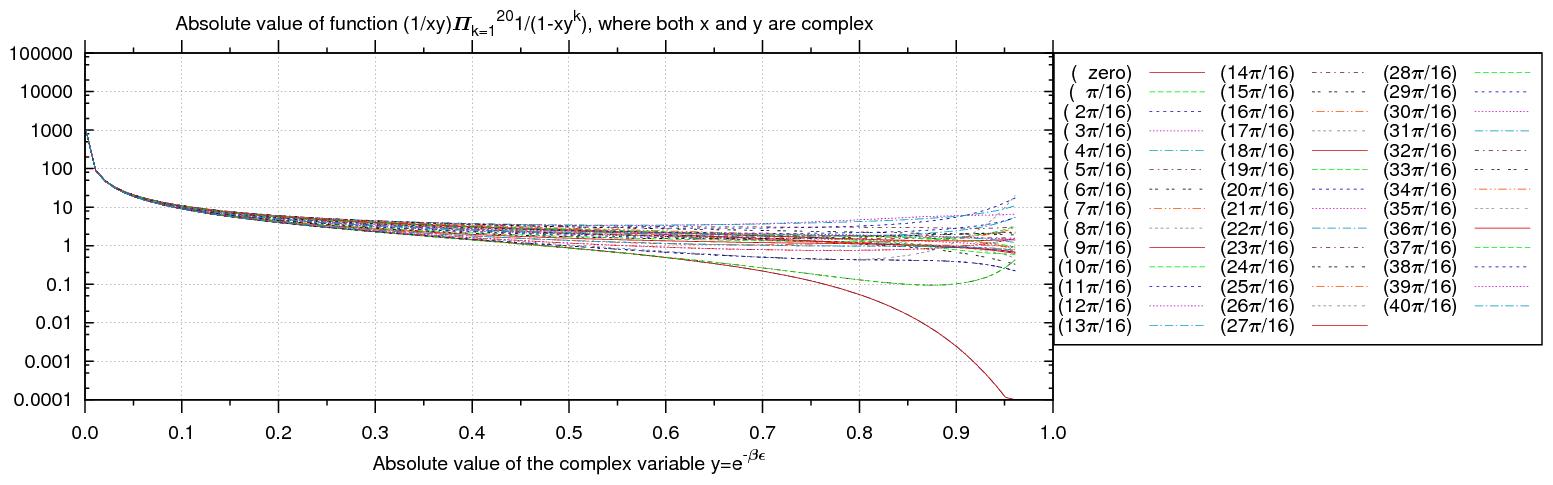}{345}{Projection of the graph of Fig. 9 on the plane containing \abs{\ya} and the
\za-axis. The various values of arg(\ya) are highlighted with different colors}{-0.6cm}{-2.5cm}

\newpage 
\makfigm{11}{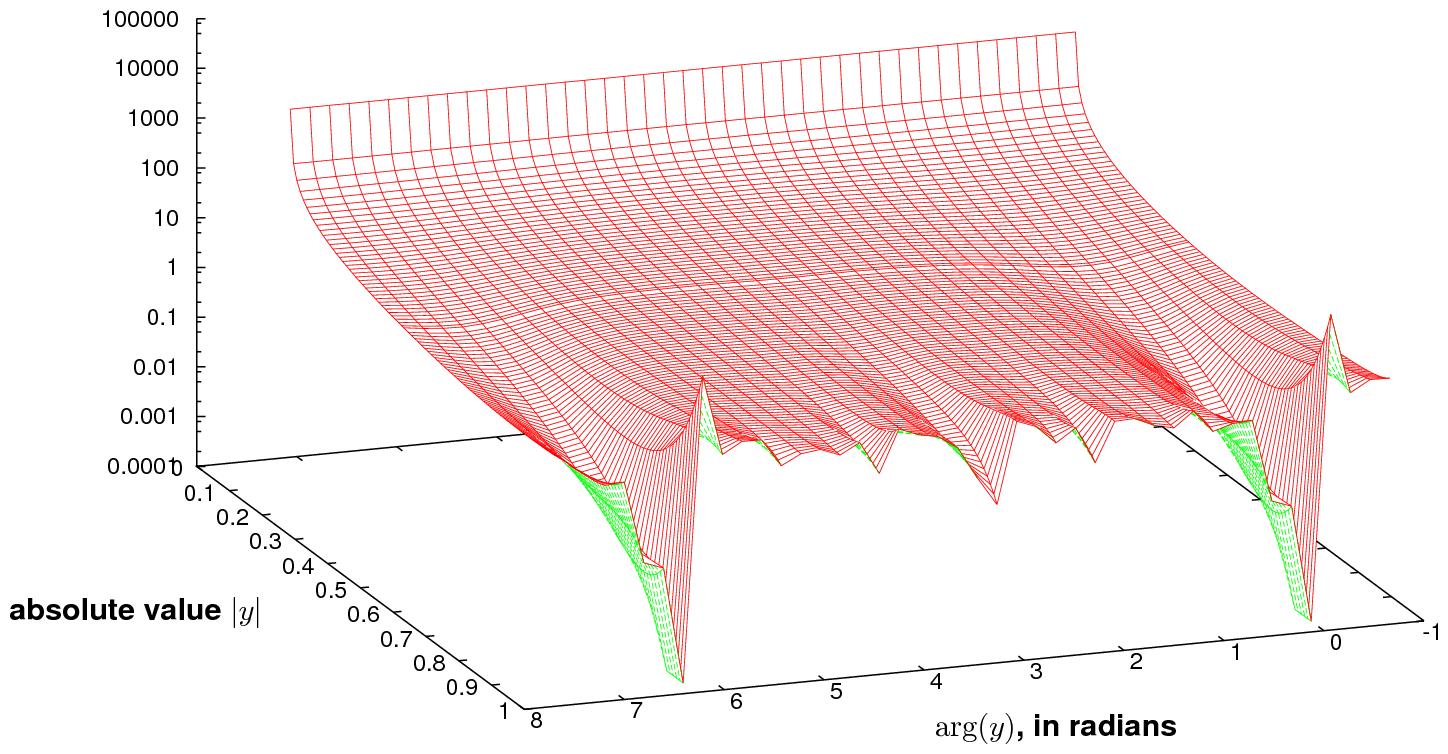}{335}{Simplified generating function for fermions \fa(\xa,\ya) of
Eq.(3.19), for variable \ya\ and fixed \xa, with arg(\xa)=7\pi/8, in the case of \mc=20
sp-states, in the fundamental state.}{-1.3cm}{-3.0cm} 
 
\makfigm{12}{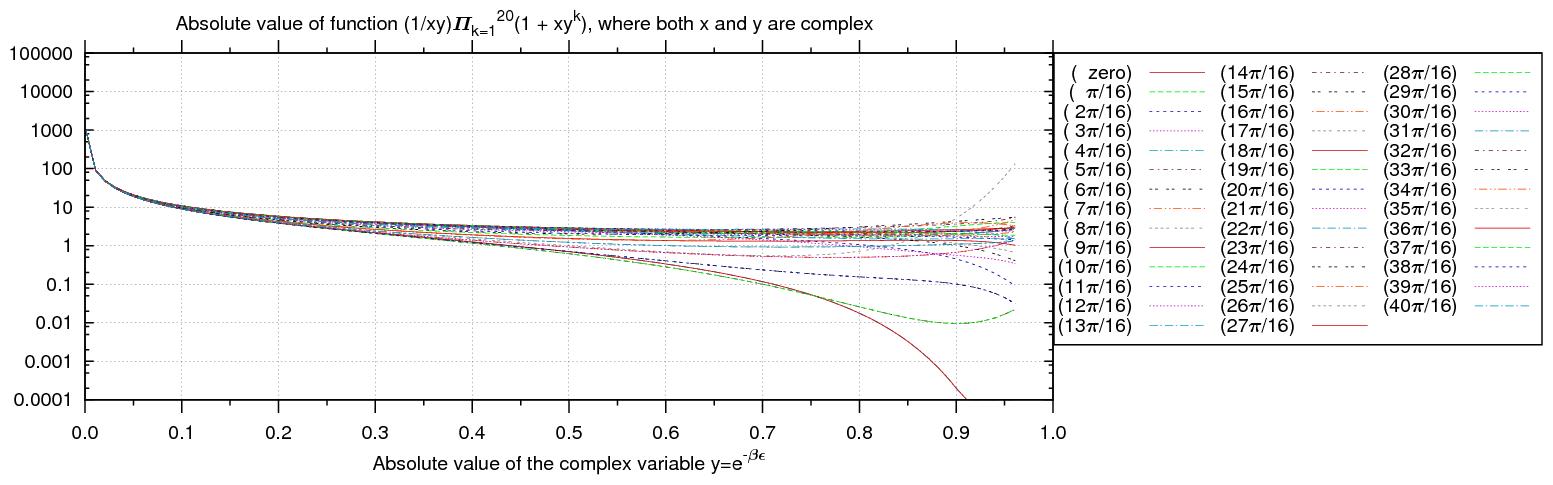}{345}
{Projection of the graph of Fig. 11 on the plane containing \abs{\ya} and the
\za-axis. The various values of arg(\ya) are highlighted with different colors}{-0.6cm}{-1.3cm}

The non null argument of \xa\ not only introduces an additional phase in the general functional dependence of
the partition function, but also \embm{changes its behavior} along the real axis from a set of points of
maximum to a set of \embm{minima}.
 
Figures \ref{figuren.13} to \ref{figuren.16} show results analogous to figures \ref{figuren.9} to
\ref{figuren.12}, but now for \mc=10 ``particles", arg(\xa)=7\pi/8 and non null excitation, \ec=4 in
arbitrary units. One important difference with respect to figures~\ref{figuren.9} to \ref{figuren.12} is the
steeper increase of the partition function for \abs{\ya} close to 0.

 \vst  \vst   \vst 
\makfigm{13}{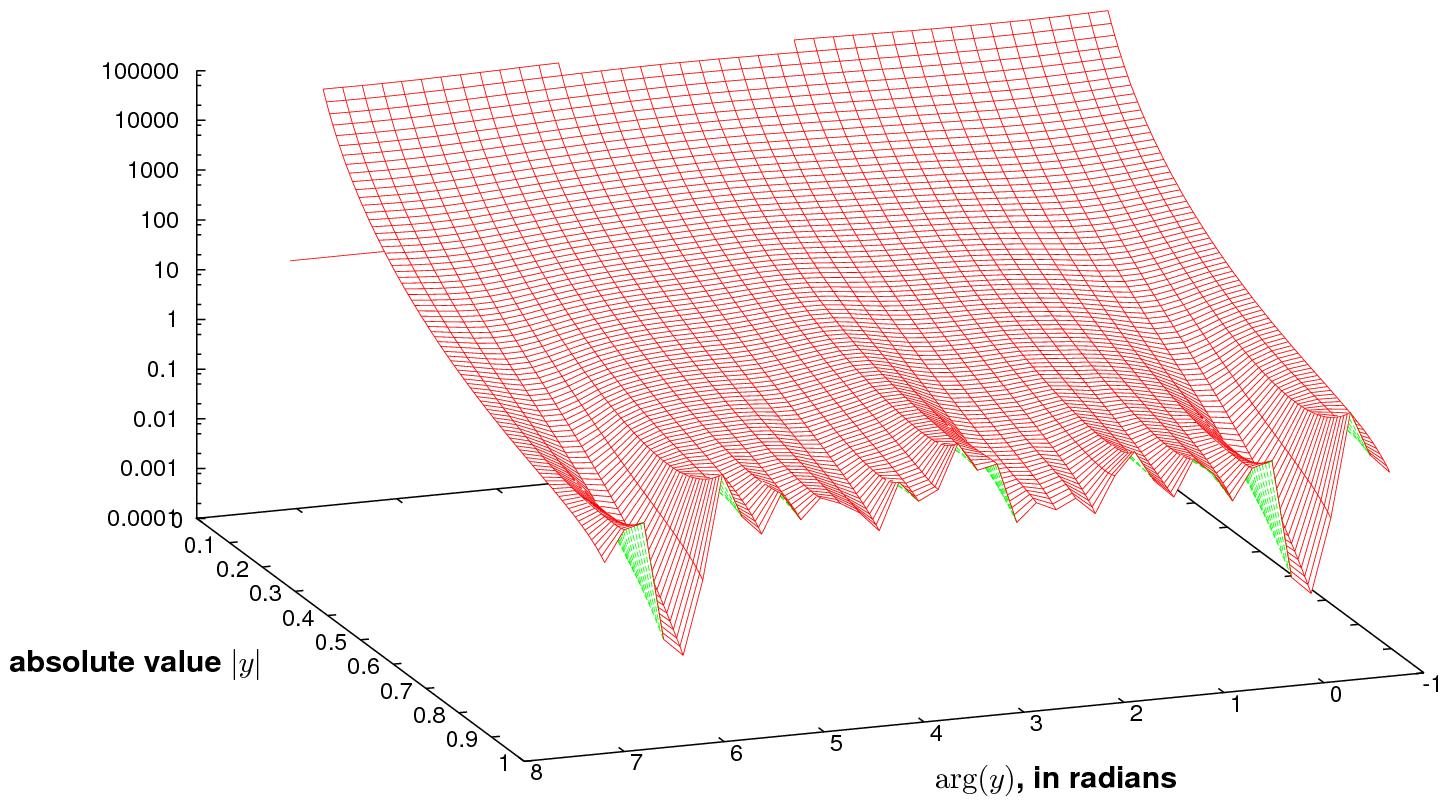}{335}{Simplified generating function for bosons \fa(\xa,\ya) of
Eq.(3.2), for variable \ya\ and fixed \xa, with arg(\xa)=7\pi/8, \mc=10 sp-states and \ec=4 in
arbitrary units.}{-1.3cm}{-2.5cm} 

\makfigm{14}{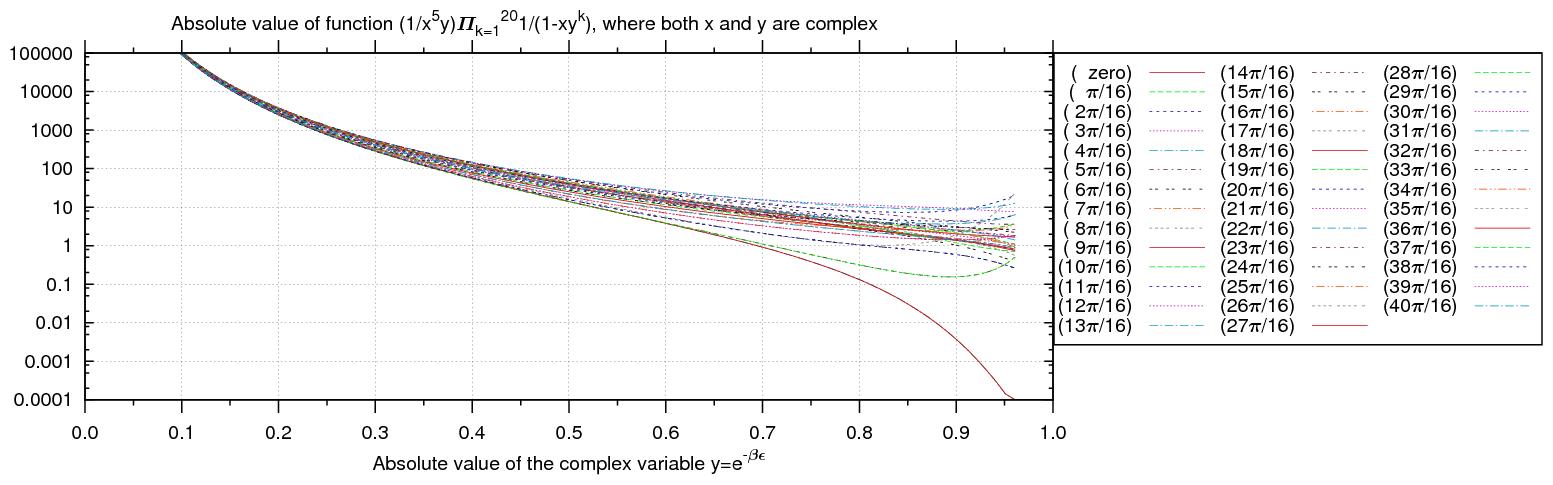}{345}{Projection of the graph of Fig. 13 on the plane containing \abs{\ya} and the
\za-axis. The various values of arg(\ya) are highlighted with different colors}{-0.6cm}{-2.5cm}

\makfigm{15}{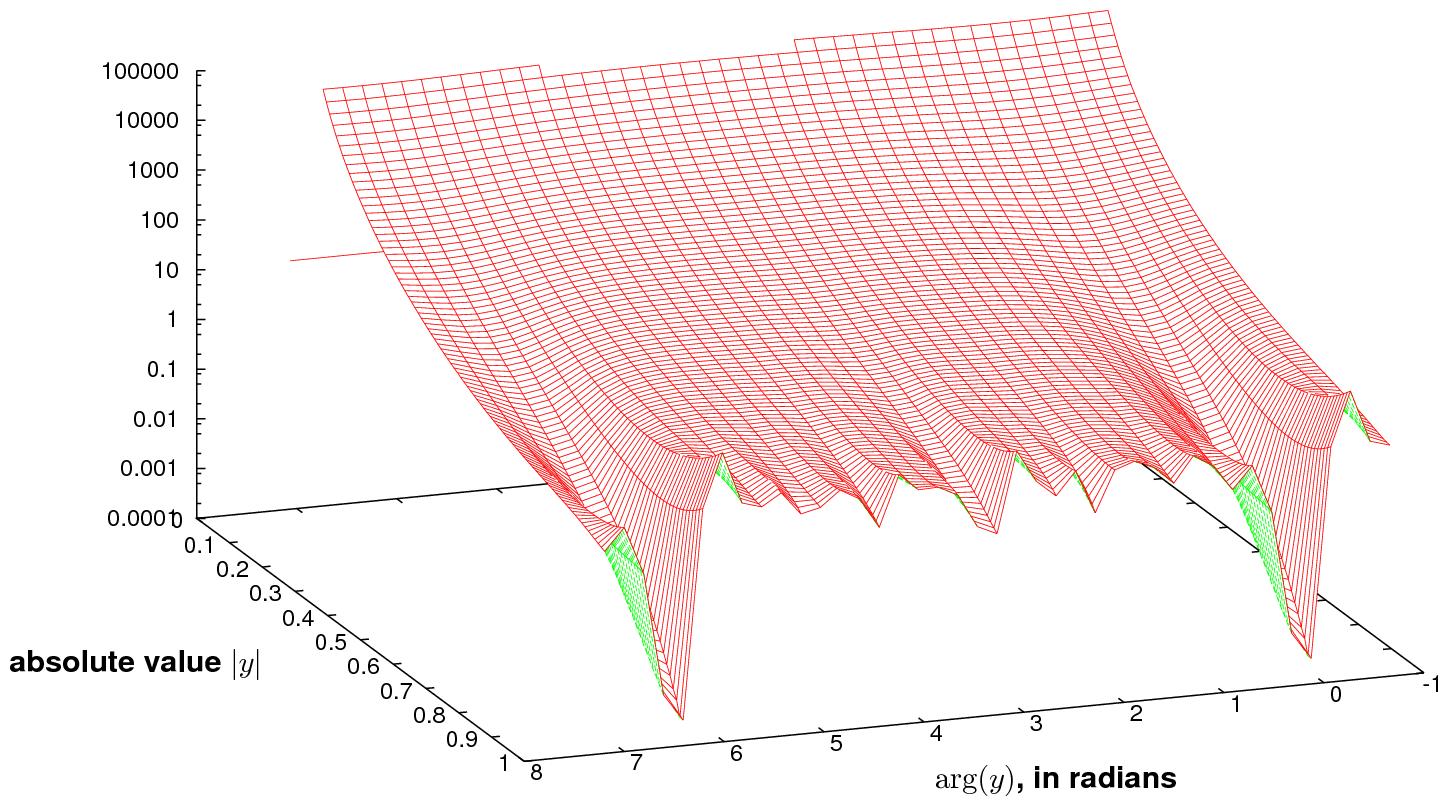}{335}{Simplified generating function for fermions \fa(\xa,\ya) of
Eq.(3.19), for variable \ya\ and fixed \xa, with arg(\xa)=7\pi/8, \mc=10 sp-states and \ec=4 in
arbitrary units.}{-1.3cm}{-3.0cm} 
  
\makfigm{16}{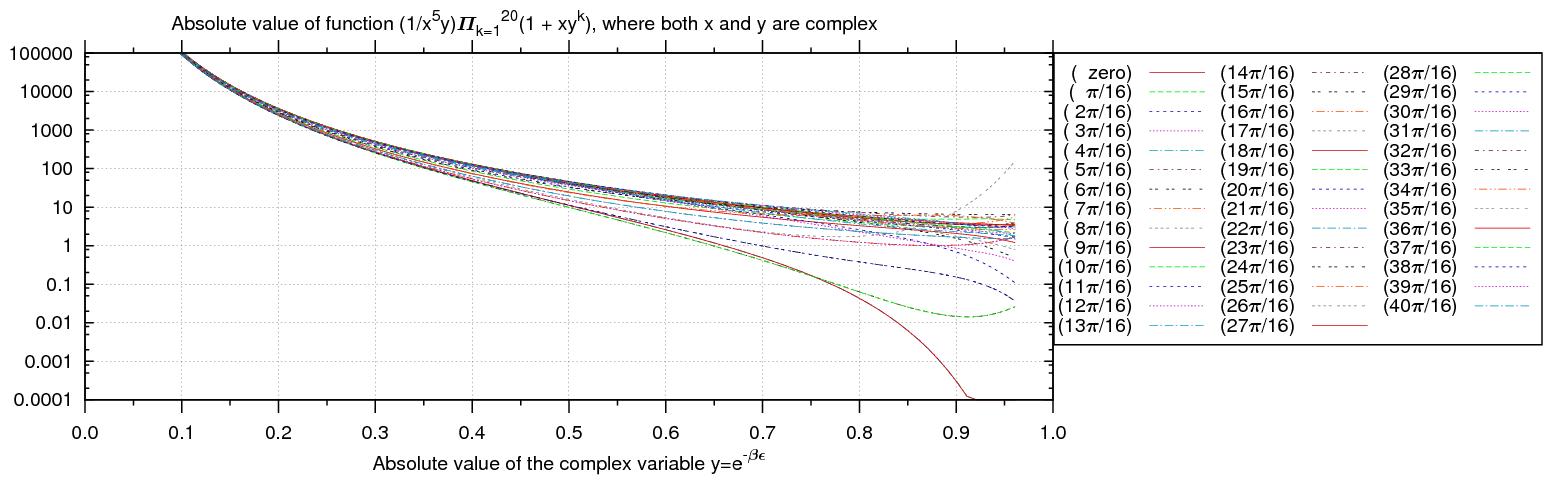}{345}{Projection of the graph of Fig. 15 on the plane containing \abs{\ya} and the
za-axis. The various values of arg(\ya) are highlighted with different colors.}{-0.6cm}{-1.3cm}

\hst Therefore, for \arg(\xa)=7\pi/8\ the reasoning of Darwin-Fowler method becomes \embm{totally flawed}.

Notice that for fermions in a nuclear system the low number of ``particles" and low nuclear excitation are
\embm{usual} assumptions in the study of pre-equilibrium dynamics,\cite{TNG} therefore in this regard the
above analysis is realistic. 

In Ref.\cite{df22} the number of PV's is implicitly assumed to be large in agreement with
the usual statistical approach. On the other hand, the above results clearly show that an increase in
\mc\ or \ec\ would only introduce more oscillations for \abs{\ya} close to 1 and increase the
steepness of the variation of the partition function for \abs{\ya} close to 0, but would not change
the main results, which show that although the maxima on the real axis of \xa\ and \ya\ may exist and be saddle
points of the partition function they cannot be considered, in general, as strong maxima unless \ec\ is very
large. In addition, if \xa\ is considered as a complex variable, the saddle points can only be defined in
certain directions of its complex plane.

Therefore, one obtains here the same conclusion as in the previous section, i. e., that the hypothesis of the
exitence of a ``strong maximum" along the direction of the contour {cannot be considered as generally
valid} as assumed in the method of Ref.\cite{df22}.

\newpage 
\secto{Final comments and conclusion}

The analysis presented in the previous sections indicates that the whole idea of using the Cauchy theorem in the
study of the partition functions of the canonical and grand canonical ensemble, for bosons or fermions,
and the consequent connection with the Laplace transform (via \emm{CAP}), as proposed by the Darwin-Fowler method
can be misleading. 

In general, the Laplace transform formalism results from the usual statistical 
interpretation of \ya\ as \eu{-\be\ep},\cite{fbgarXiv1} where \be=1/\kapp\tc, \kapp\ is Boltzman constant, \tc\ is the
nuclear temperature and \ep\ is the average interspacing among single particle levels, complemented by the
hypothesis that \ep\ is infinitesimally small in accordance with \emm{CAP}. 
Then, one can rewrite for example \ree{2.18} as
\bm{ f(x,y) = \sums{A,\uc} \cald(\ac,\uc)\xau{A}\yau{\uc/\ep} \app 
 \sums{\ac} \intsu{\ecs{\min}}{\ecs{\max}} \om(\ac,\uc) \xau{A} \eu{-\be\uc}\da\uc 
\;,}
where the nuclear density \om(\ac,\uc) and degeneracy \cald(\ac,\uc) are related by 
\bn{\ome\brac{\ac,\uc,\mc}\approx \cald(\ac,\uc,\mc)/\de\uc \;,} 
and
\bm{ \de\uc = \uc-\ucs{\prev}  \;, }  
where \ucs{\prev} is the highest (discrete) nuclear level energy smaller than \uc. Therefore, with the
approximate replacement of (\ecs{\min},\ecs{\max}) by (0,\infty), which is reasonable if \uc\ is for example
the nuclear excitation, \fa(\xa,\ya) becomes the Laplace transform of \om(\ac,\uc).

The approximations involved here are not drastic having in sight the usually high density of nuclear
levels per MeV observed experimentally,\cite{TNG} which makes the use of \emm{CAP}, in general, a very
reasonable approach. Then, because the connection of approximated expressions involving the Cauchy's theorem,
like \ree{2.6} or \ree{2.11}, with the inverse Laplace transform is also immediate under
\emm{CAP}\cite{fbgarXiv1} the various formalisms become intimmately related and tend to be used together,
which may be a source of errors if one of these approximations is at least partially incorrect. 

For example, in Ref.\cite{fbgarXiv1} it was shown that the direct use of the Laplace transform may lead to
inconsistencies in the description of transitions in which a given sp-state is destroyed and subsequently
re-created. 

In this case, the Laplace transform is not able to describe the details of the variation of the
set of available states from initial to intermediary and final stages, due to the approximations
involved in the definition of the nuclear density. In this case, the imprecise definition of the 
density may give a wrong result for the moments of the transition if not seconded by an independent
analysis of the involved microscopic processes and their influence on the definition of the available
states at each stage. 

Although this problem with the Laplace transform is essentially independent of the Darwin-Fowler
method it would not have appeared if the method was not employed in the first place, as the Laplace
transform is not a necessary tool for the description of pre-equilibrium nuclear systems. 

One may ask, if the method of Ref.\cite{df22} has this essential inconsistency and it is not so general
as it is usually supposed to be then why does it work so well in many applications? A possible answer for this
may be found in the details of the method itself and the meaning of the functions that the method is
supposed to be applied to.

For a given analytical function \ga\of{\ya} that is non zero at \ya\ the following relation is always valid,
\bm{ \gap\of{\ya} = \frac{\da}{\da\ya} \brak{\log(\ga\of{\ya})} \ga\of{\ya} \;. }
In the case of an ensemble of bosons or fermions \ga\of{\ya} is the partition function of the ensemble,
\fa\of{\ya}, divided by terms like \yau{\nc+1}, where the total energy is \ec=\nc\ep. Therefore, the minimum
is given by the following equation
\bm{ \frac{\da}{\da\ya} \brak{\log\of{\ga\of{\ya}}} = \frac{\da}{\da\ya} \brak{\log\of{\fa\of{\ya}} -
(\nc+1)\log\of{\ya}} = 0  \;,}
and by making the usual connection between the partition function and the thermodynamic potential, i. e.
\log\of{\fa\of{\ya}}= -\be\omc, where \omc\ is given by\cite{Fetter} 
\bm{ \omc = \ec - \tc\sc - \mu\mc = \nc\ep - \tc\sc - \mu\mc \;, }
results 
\bm{ -\be\omc = (\nc+1) \log\of{\ya} + \ea \;,}  
where \ea\ is a constant of integration that is not a function of \ec, therefore one can take \ea\ as equal
to \be(\tc\sc+\mu\mc) to obtain
\bm{ -\be\ec = (\nc+1)\log\of{\ya} \app \nc\log\of{\ya} \mbox{\hstps ; if \nc\ is very large, } \;}
which is the usual thermodynamic interpretation of the statistical parameter \ya. 

Therefore, the condition of minimum in \ree{5.4} can be interpreted as an approximate expression of the usual
relation between the thermodynamic potential and the elementary energy associated with the single particle
quantum levels,
\bm{ \frac{\del\omc}{\del\ep} = -\be\nc \;,}  
showing that the entire use of the Cauchy theorem in connection with expressions involving the partition
functions, for an ensemble of either fermions or bosons, can be seen as a tricky reformulation to
obtain the usual thermodynamic expressions from the microscopic description defined by the partition
functions. 

In other words, the above reasoning shows that the condition of focusing on the minimum of the
integrand instead of the entire integral defined by the Cauchy theorem would work well in the case of
partition functions even if the integrands do not define a saddle point with the characteristics
assumed in Ref.\cite{df22}.  

Therefore, the use of the Darwin-Fowler method can be considered as a
convenient tool for many applications, but cannot be taken as the foundation of the analysis of
ensembles of fermions or bosons, as suggested by Ref.\cite{b68}. 

Consequently, the use of \embm{direct algebraic approaches}, as the one presented in Ref.\cite{fbgarXiv1}
for the analysis of pre-equilibrium dynamics, can be more appropriate for a microscopic description of the
Shell Model, even in the \emm{CAP} limit and either for fermionic or bosonic systems, unless the total energy
\ec\ is very large. 

\vst\vst  \vst\vst \vst\vst 
\mbm{\LZ \hst Acknowledgments}

The author is pleased to acknowledge the friendship and support of the colleagues IEAv/DCTA during the
realization of this work.

\newpage      

\addtocounter{sectionp}{1}
\setcounter{section}{\value{sectionp}}

\end{document}